\documentclass[prl,aps,amssymb,amsmath]{article}
\usepackage{amssymb,amsmath}
\usepackage{chicago-bibstyle}  % use this style if you don't use BibTeX.
\usepackage{latexsym}
\newtheorem{twr}{Theorem}
\newtheorem{lem}{Lemma}

\begin{document}
\title{{\bf Multiplication in Newton's \emph{Principia}}}
\author{Jaros{\l}aw Wawrzycki \footnote{Electronic address: jaroslaw.wawrzycki@wp.pl} \\
Institute for the History of Science, \\
Polish Academy of Sciences, Warsaw\\
\\
\ \ \ \ \ \ \ \ \ \ \ \ \ \ \ \ \ \ \ \ \ \ \ \ \ \ \ \ \ \ \ \ \ \ \ \ 
 \ \ \ \ \ \   \emph{To the Memory of Stefan Banach}}
\maketitle
\newcommand{\ud}{\mathrm{d}}

\vspace{1cm}

\begin{abstract}
Newton in his \emph{Principia} gives an ingenious generalization of the Hellenistic
theory of ratios and inspired experimentally gives a tensor-like definition of multiplication of quantities measured with his ratios. An extraordinary feature of his definition is generality: namely his definition a priori allows non commutativity of multiplication of measured quantities, which may give a non-trivial linkage to experimental facts subject to quantum mechanics discovered some two hundred years later.
Mathematical scheme he introduces with this ingenious definition
is closely related to the contemporary approach in spectral geometry.
His definition reveals in particular that commutativity of the multiplication of quantities with physical dimension has the status of experimental assumption
and does not have to be fulfilled in reality, although neither the mathematical 
tools nor experimental evidence could allow Newton to carry out
the case when the multiplication is noncommutative. We present a detailed analysis of his definition as well as a linkage to the theory of representations of algebras with involution (and with normal forms of von Neumann algebras and Jordan Banach algebras).  
\end{abstract}

\section{Introduction}
\label{intro}

In his extraordinary epoch-making book \emph{Philosophiae naturalis principia mathematica} \cite{Newton}, commonly known under the
short name \emph{Principia}, Newton developed axiomatical mechanics of motion, and gave extraordinarily effective computational tools
allowing to compute enormous variety of quantities verifiable experimentally, and uncovered connections of
what was seemed unconnected before. Here we concentrate on the theory of \emph{proportions} or rather on 
the Newton's usage of the Eudoxus theory of \emph{ratios}. Newton gives a revolutionary view on the notion of ratio and proposes to extend Eudoxus' theory so as to embrace his fluents, fluxions, 
and his momenta (geometric representors of differentials) thus using 
ratios of not only such quantities as rectangles, 
their sides, and so on but also he uses \emph{ratios of fluents}, or of fluxions, or momenta. 
Although Newton does not axiomatize the \emph{proportion} (or \emph{ratio}) of such general quantities as fluents explicitly nor gives any axiomatic or descriptive presentations of it, 
nonetheless he gives in the \emph{Principia} a very consequent usage of \emph{ratio, quantity} and \emph{number}, and proceeding by hints and examples gives a very interesting programme for developing analysis and geometry which was ahead of his times but which fits very well with the contemporary mathematics and physics and in fact lies at the very front of nowadays mathematics. Here we extend slightly Newton's suggestion, and try to give it more explicit shape, although the original Newton's presentation is an art \emph{en soi}.  Hidden among research into celestial mechanics, this ingenious programme has hardly been drawn to the attention of mathematicians. To my knowledge only one of the greatest giants of mathematics, Bernhard Riemann and Richard Dedekind seemed to be inspired by this programme, but their works inspired by it were themselves hardly accessible to the mathematical community. We therefore think that our accout is well motivated.            

Newton was very brief in presenting mathematical tools used in 
\emph{Principia} which he founded
himself, but his short hints was ingeniously right, and as the history of mathematics revealed they pointed decisive constructions allowing us to
bypass the hardest difficulties. For example at his time
there were some paradoxes in analysis, and among them the so called
\emph{paradox of indivisibles}, which undermined the so called \emph{method of indivisibles} commonly used at Newton's times. Newton pointed out that the Archimedes method supported on the Eudoxus theory of ratios may be extended and used to solve the difficulties encountered in analysis, or, what is the same thing, that the Eudoxus theory of ratios may be extended allowing us to make ratios of fluents, fluxions, \dots. This may seem very
unexpected for us today, because this idea requires a study of function-like
spaces with non-trivial order structure and linear topology -- an approach
developed systematically only at the beginning of the previous century mainly by Banach. This is indeed the case however, because Newton founded his mathematics on ratios and proportions treating the ratios of geometric representors of fluents (functions) as the analogue of the ordinary ratios of Eudoxus theory. We must remember that only after Dirichlet we have the Cantor-like definition of function at our disposal. Newton did differently. Moreover, he treated in the same way the theory of algebraic and real analytic functions -- representors of some geometric quantities --  treating them as algebraic elements composing whole spaces\footnote{Better: composing fields, i.e. with the inversion for every non-zero element.}, which may be multiplied and added and which may be resolved into rapidly converging power series.\footnote{Newton in his writings explicitly points out that the Archimedean quadrature of parabola gives a strict proof that the geometric series is convergent (Barrow's notice). On the same footing he argues that his power series are convergent although no explicit proof can be found in his mathematical writings.} This idea was afterwards undertaken by Riemann in his theory of complex analytic functions and especially by Dedekind and Weber in their algebraic theory of Riemann
analytic functions. Newton gives some examples of this idea generalizing Eudoxous ratios and hidden it among the mathematical Lemmas of the \emph{Principia}, and among the consequent usage of the term \emph{ratio} and of the \emph{geometric quantities} for which the ratio he considers throughout the whole \emph{Principia}.\footnote{The modifications differing the first three authorized Latin editions are not important for the Newton's treatment of Eudoxus ratios discussed here.} From this one can read of the general picture, which we describe in intuitive terms first, and develop in the following Sections. 
 
Thus Newton distinguishes, after the Hellenistic mathematicians, between the \emph{geometric quantities}\footnote{Hermann Weyl calls them 
simply \emph{magnitudes} in his philosophical writings \cite{W1} and \cite{W2}.} and \emph{ratios} which they may compose. Ratios are understood by him as operations acting on geometric quantities with magnitude and are think of as of some \emph{algebraic quantities} which may be multiplied in the sense of composition of operations. And just
as the operations of dressing do not commute, also here the multiplication a priori may be non-commutative (dressing first the dress shirt and then the coat is not the same as dressing first the coat and then the dress shirt). The geometric quantities which compose ratios are never multiplied by Newton in the same sense as ratios. This distinction is very consequent throughout the whole \emph{Principia} with
(almost) no exceptions, and is reflected not only in computational manipulations but also in the notation. For example Newton never writes $AB^3$ or $AB^2$ for the side $AB$ (or segment $AB$) or for any geometric quantities $AB$ which may compose a ratio. The notation $X^3$ or $X^2$ is reserved for algebraic quantities, such as ratios $ X = \frac{AB}{CD}$, acting on the geometric quantities $CD$. For geometric quantities he writes instead $AB \times AB \times AB$ or $AB \times AB$, and interprets as a solid or rectangle
if $AB$ is a segment of a line. That rectangle may represent multiplication of respective
ratios requires a proof, and may again be read of from the \emph{Principia} although it is encoded between the lines (we give details in the following Sections). Generally the geometric quantities are quantities with magnitude, say a dimension -- as we call it today, and may be measured in the sense of (generalized) theory of ratios. Such quantities cannot be multiplied as dimension-less algebraic
quantities such as ratios. Newton notices it, and generalizes the construction 
of the rectangle $AB \times AB$ introducing something we call it today a 
\emph{tensor product}, with a universal bilinear $\cdot \times \cdot$ function, and calls \emph{conjunct dependence}. Then proposes $\times$ to represent multiplication of his generalized ratios. Because the bilinear tensor function $\times$ may a priori be non-symmetric (just as the ratio may a priori be non-commutative) if the space of geometric quantities is higher dimensional, Newton created general mathematical frames embracing (or allowing to) non-commutative character of observable quantities.
Although Newton created such general mathematical frames there where both no observable 
indications and no mathematical tools (functional analysis) needed to investigate
them in full generality. This ingeniously general mathematical frame presented already in
\emph{Principia} has been forgotten, both by physicists and by mathematicians.
Only at the beginning of 20th century had the experimental discoveries around
quantum mechanics revived a knew mathematical structure (von Neumann and C*-algebras) 
and approach to geometry (non-commutative geometry),
whose general ideas could be read of from Newton's \emph{Principia}.
Here we present a closer look at the generalized theory of ratios as suggested by 
Newton and its connection to the \emph{conjunct dependence}-tensor like character
structure of measured geometric quantities.

\section{Eudoxus' theory of ratios}

Let us remind briefly the rudiments of the Eudoxus theory of ratios
in order to understand better the idea of Newton and afterwards we
get over to Newton's generalization.
Of course information of it comes from Euclid \emph{Elements}
and the other Hellenistic works, especially those of Archimedes and Apollonius -- sources studied by Newton. Eudoxus' theory of ratios
provides us a transition from \emph{counting} to \emph{measuring} of geometric quantities, such as segments on a line. Within the domain of the geometric quantities there is defined a \emph{congruence} relation $=$. The geometric quantities $\boldsymbol{a}, \boldsymbol{b}, \dots$ the ratios of which are to be considered may be added together and are endowed with a magnitude which allows us to introduce an order relation $<$ between them. Addition $+$ and order $<$ are compatible: if $\boldsymbol{a} < \boldsymbol{c}$, then $\boldsymbol{a} + \boldsymbol{b} < \boldsymbol{c}+\boldsymbol{b}$. From a geometric quantity 
(say a segment\footnote{But Eudoxus, Euclid or Archimedes consider various geometric quantities. For example Archimedes considers plane figures, with the equivalence relation $=$ meaning the equivalence by a triangle decomposition: two figures are equivalent whenever may be decomposed into congruent triangles. Addition $+$ and $<$ are self-evident: a figure 
$\boldsymbol{a}$ is a sum $\boldsymbol{b} + \boldsymbol{c}$ of 
$\boldsymbol{b}$ and $\boldsymbol{c}$ whenever there exist triangle decompositions $D_{\boldsymbol{a}}, D_{\boldsymbol{b}}, D_{\boldsymbol{c}}$ of $\boldsymbol{a}, \boldsymbol{b}, \boldsymbol{c}$ such
that $D_{\boldsymbol{a}}$ is a set theoretical sum of
$D_{\boldsymbol{b}}$ and $D_{\boldsymbol{c}}$. For the set theoretical 
culture of Archimedes' writings compare the investigations of Reviel Netz
and his co-workers: \cite{netz1}, \cite{netz2} and \cite{noel}, or
compare the comments on actual infinity in Chap. 2 of their popular book 
\cite{Netz} (unfortunately their new critical edition 
of the Heiberg-Archimedes palimpsest is still in preparation). 
Of course under the assumption of Archimedes-Eudoxus
postulate and order completeness all such spaces of congruence classes of geometric objects are mutually isomorphic as to their algebra-order structure and are all isomorphic to the field of real numbers (Dedekind).}) $\boldsymbol{a}$ one can compose the quantity 
$2\boldsymbol{a}$, $3\boldsymbol{a}$, \dots or  $n\boldsymbol{a}$, by forming
the sum $\boldsymbol{a} + \boldsymbol{a}$, $\boldsymbol{a} +\boldsymbol{a}
+\boldsymbol{a}$ \dots or $\boldsymbol{a} + \dots + \boldsymbol{a}$ with
2, 3, \dots or $n$ terms; which brings out the connection
between counting and measuring \cite{W1}. Slightly more formally we may
define the iteration as follows:
\[
\begin{array}{l}
\alpha) \,\, 1 \boldsymbol{a} = \boldsymbol{a},\\
\beta) \,\, \textrm{If $n$ is a natural number, then $(n+1)\boldsymbol{a}$
results} \\
\textrm{from $n\boldsymbol{a}$ by the following formula}\\
(n+1)\boldsymbol{a} = (n\boldsymbol{a}) + \boldsymbol{a}.
\end{array}
\]        

The assumption that every geometric quantity $\boldsymbol{a}$ may be measured by every other $\boldsymbol{b}$ Eudoxus, Euclid and Archimedes expressed as follows:

\begin{enumerate}
\item[]
{\bf Archimedes-Eudoxus postulate} \, \emph{For every two geometric quantities $\boldsymbol{a}$ and $\boldsymbol{b}$ there exists a natural number $n$ such that $n\boldsymbol{a} > \boldsymbol{b}$.}
\end{enumerate}
The equality of ratios is expressed by Euclid as follows:
\begin{enumerate}
\item[]
{\bf Eudoxus-Euclid definition} \, \emph{Two ratios, $\boldsymbol{a':a}$
and $\boldsymbol{c':c}$, are equal to each other if for arbitrary natural
numbers $m$ and $n$}
\[
\begin{array}{l}
\textrm{(I)} \,\,\, \textrm{\emph{from $n\boldsymbol{a'} < m \boldsymbol{a}$ it follows $n\boldsymbol{c'} < m \boldsymbol{c}$}} \\
\textrm{\emph{and}} \\
\textrm{(II)} \,\,\, \textrm{\emph{from $n\boldsymbol{a'} = 
m \boldsymbol{a}$ it follows $n\boldsymbol{c'} = m \boldsymbol{c}$}} \\
\textrm{\emph{and}} \\
\textrm{(III)} \,\,\, \textrm{\emph{from $n\boldsymbol{a'} > m \boldsymbol{a}$ it follows $n\boldsymbol{c'} > m \boldsymbol{c}$}}. \\

\end{array}
\]   
\end{enumerate}
In the original works of Archimedes or Euclid, which place the Eudoxus ratios
among the wider context of geometry (the same state of affairs we find
in \emph{Principia}), there are some postulates
concerning the geometric quantities used without mention, which nevertheless are important for the Eudoxus method of measuring by ratios. For example
we assume that the operation of iterate addition admits of a unique inversion, which we could suggestively call partition: given a geometric
quantity $\boldsymbol{a}$ and a natural number $n$ there exist one and (up to the congruence $=$) only one geometric quantity $\boldsymbol{x}$ such
that $n\boldsymbol{x} = \boldsymbol{a}$; denoted by $\boldsymbol{a}/n$. 
Thus iteration and partition are viewed as operators acting in the domain of geometric quantities, in particular iteration may be combined with partition. In particular $n\boldsymbol{a}/m$ serves as a composite operation of iteration $(n,\boldsymbol{a}) \mapsto n\boldsymbol{a}$ and partition
$(m,\boldsymbol{a}) \mapsto\boldsymbol{a}/m$, and is called action of $n/m$ on $\boldsymbol{a}$. The fractional symbol $n/m$ denotes the composite operation, and thus two such operations are equal whenever they are equal as operators acting in the domain of geometric quantities, i.e. whenever lead to the same result, no matter to what geometric quantity 
$\boldsymbol{a}$ they are being applied. \emph{Multiplication} of fractions is performed by carrying out one after another the operations which are to be multiplied. In practice it is sufficient to investigate the behaviour of fractional operations on the sub domain $n\boldsymbol{a}$, with $n$ ranging over the natural numbers. The fact that 
$n\boldsymbol{x} = m\boldsymbol{y}$ cannot always for given 
$\boldsymbol{x}$ of the form $n\boldsymbol{a}$ be solved within the
sub domain of iterated $\boldsymbol{a}$ does not matter here. Using the properties of multiplication within the natural numbers we easily see
that $n/m$ and $n'/m'$ are equal iff 
\[
nm' = mn'.
\]     
It is unnecessary thus to introduce special fractions for each domain of geometric quantities, as we can characterize them in representation-free
manner. This fundamental observation was already present in the Sumerian
mathematical culture, and the same way of thinking is present in the 
\emph{Principia}. Newton's way of thinking about more general fractions,
namely Eudoxus ratios is the same. He treats them as operations acting
on the geometric quantities. The only difference is that because there were no purely arithmetic construction of Eudoxus ratios at his disposal independent of representation (Dedekind) Newton proposes to consider both: the algebra of ratios together with the space of geometric objects acted on by the ratios; i.e. he proposes to proceed after successful Archimedes  \emph{method of exhaustion}, compare e.g. Newton's comments
in the Scholium ending the first Section of the first Book of \emph{Principia}. Here is the crucial difference between the 19th century
mathematics and that of Newton. For Newton (and for Eudoxus,
Archimedes or Euclid) it is up to the postulates (axioms) of geometry to tell
us what geometric object ratios do exist. The algebra of ratios describing geometry was constructed from geometry, not conversely. It is only after Dedekind's arithmetization of real number that the direction can be inverted. Newton (and Barrow), e.g. in the Scholium mentioned to above recognized 
that the problem with indivisibles may be resolved if one appeals to the 
continuity just as Archimedes did in his quadrature of parabola: although
we cannot express the completeness in terms of ordinary fractions, we can nevertheless express it as Eudoxus ratios.

Thus Newton, when suggested to extend the Eudoxus method of ratios in order to way out of the ''labyrinth of the continuum''(Leibniz's known expression)
proposes to consider ratios of geometric representors of his fluents, their
fluxions (derivative) or momenta (differentials). \emph{Principia} are written geometrically. In the first Lemmas the reader will find purely geometric representation of (as we would say today) the Taylor formula with
the Peano remainder together with the proof that the ratio of the remainder  to the first terms goes to zero whenever we tend to the origin of the development. Generally all computations are given geometric expression, and Newton, although proceeds by (an enormous variety of) examples
does not give any axiomatic formulation of his idea. It is rather evident
that he hopes the geometric context to be sufficiently reach for the program to be realized. He believes that the adequate order relation $<$ and addition $+$ should be within a constructive reach sufficient for extension of the method of ratios. In particular in the space of geometric objects there should exists distinguished elements $\boldsymbol{a}$ (order units, as we call them today: order unit measures every other geometric quantity, i.e. for any geometric quantity $\boldsymbol{b}$ there is a natural $n$ such that $n\boldsymbol{a} > \boldsymbol{b}$) such that at least for some geometric objects $\boldsymbol{b}$, the pair $\boldsymbol{a}$, $\boldsymbol{b}$ defines a ratio $\boldsymbol{b:a}$. However the strong Archimedes-Eudoxus postulate has to be weakened, so that not every geometric quantity may be measured by every other, in particular $\boldsymbol{b}$ not necessarily measures $\boldsymbol{a}$, although they may compose a well defined ratio 
$\boldsymbol{b:a}$. The ratios -- viewed as operations acting on geometric quantities -- may be added and multiplied (in the sense of composing operations) if for any four such quantities $\boldsymbol{a},\boldsymbol{b}, \boldsymbol{c}, \boldsymbol{d}$, such that the first two and the last two compose a ratio: $\boldsymbol{a:b}$ and 
$\boldsymbol{c:d}$, there exist other geometric quantities respectively
\[  
(1) \,\, \begin{array}{c}
\boldsymbol{a'},\boldsymbol{b'},\boldsymbol{c'} \,\,\,
\textrm{such that} \\
\boldsymbol{a':b'} = \boldsymbol{a:b} \,\, \textrm{and} \,\,
\boldsymbol{c':b'} = \boldsymbol{c:d}   
\end{array} \,\,\,
(2) \,\, \begin{array}{c}
\boldsymbol{a''},\boldsymbol{b''},\boldsymbol{d''} \,\,\,
\textrm{such that} \\
\boldsymbol{a'':b''} = \boldsymbol{a:b} \,\, \textrm{and} \,\,
\boldsymbol{b'':d''} = \boldsymbol{c:d}  
\end{array}
\]
\[
(3) \,\, \begin{array}{c}
\boldsymbol{a''},\boldsymbol{b''},\boldsymbol{c''} \,\,\,
\textrm{such that} \\
\boldsymbol{a''':b'''} = \boldsymbol{a:b} \,\, \textrm{and} \,\,
\boldsymbol{c''':a'''} = \boldsymbol{c:d}.  
\end{array}
\]   
We have then
\[
(1) \,\, \begin{array}{c}
\boldsymbol{a:b} + \boldsymbol{c:d} = \boldsymbol{a':b'} 
+ \boldsymbol{c':b'} = (\boldsymbol{a'+ c'})\boldsymbol{:b'}, \,\,\, \\
(\boldsymbol{c:d}) + (\boldsymbol{a:b}) = \boldsymbol{c':b'} 
+ \boldsymbol{a':b'} = (\boldsymbol{c'+ b'})\boldsymbol{:b'}, \\
\end{array}
\]  
\[
\begin{array}{cc}
(2) & \, (\boldsymbol{a:b}) \cdot (\boldsymbol{c:d}) = (\boldsymbol{a'':b''}) \cdot (\boldsymbol{b'':d''}) = \boldsymbol{a'':d''},\\
(3) & \, (\boldsymbol{c:d}) \cdot (\boldsymbol{a:b}) 
= (\boldsymbol{c''':a'''}) \cdot (\boldsymbol{a''':b'''}) 
= \boldsymbol{c''':b'''},   
\end{array}
\] 
In particular in the special case when the Archimedes-Eudoxus postulate is assumed to be valid, then from the Eudoxus-Euclid definition of equality of ratios (and from continuity, i.e. order completeness) it follows that the conditions (1)-(3) are fulfilled and the ratios may be added and multiplied. Note however that commutativity of multiplication must be proved and it is not a trivial consequence of initial assumptions in this approach. In fact the commutativity of multiplication of ratios is a theorem in the Eudoxus theory, and in the Hellenistic terminology (Archimedes) the low of commutativity is called \emph{ex aequali in perturbet proportion}\footnote{In translation of Heath, compare his famous comments to Euclid \emph{Elements} attached to his translation of \emph{Elements}: \cite{elements}. Newton uses the phrase: 
\emph{ex aequo perturbate} or \emph{vicissim} or \emph{alternando}, while Motte in his translation of \emph{Principa} uses \emph{alternando} or \emph{permutation}.}: if $\boldsymbol{a:b} = \boldsymbol{b':c'}$ and 
$\boldsymbol{b:c} = \boldsymbol{a':b'}$, then  
$\boldsymbol{a:c} = \boldsymbol{a':c'}$ and it is formulated as Proposition 23 in Book V of Euclid \emph{Elements} \cite{elements}. 

\vspace{0.5cm}

{\bf Remark} \,\,\, Note that in fact Archimedes uses the definition of equality of ratios mentioned to above
and used the (order) completeness of the domain of geometric objects. Dedekind noted that the ratio, as follows from the Eudoxus-Euclid definition, divides the space of fractions into three classes: I, II, III. The II class may contain one element at most, and every fraction of I is less then any of II (if not empty) and of III. Joining I and II (or II with III) we obtain a Dedekind's cut. And in geometry we postulate (Dedekind's axiom) the existence of that geometric object which stands to the given unit object in the ratio determined arithmetically by the cut. Thus joining of the classes 
II and III corresponds to the following reformulation of Eudoxus-Euclid definition:
\begin{enumerate}
\item[]
{\bf Another variant of Eudoxus-Euclid definition} \, \emph{Two ratios, 
$\boldsymbol{a':a}$ and $\boldsymbol{c':c}$, are equal to each other if for arbitrary natural
numbers $m$ and $n$}
\[
\begin{array}{l}
\textrm{(I)} \,\,\, \textrm{\emph{from $n\boldsymbol{a'} < m \boldsymbol{a}$ it follows $n\boldsymbol{c'} < m \boldsymbol{c}$}} \\
\textrm{\emph{and}} \\
\textrm{(II)} \,\,\, \textrm{\emph{from $n\boldsymbol{a'} \nless 
m \boldsymbol{a}$ it follows $n\boldsymbol{c'} \nless m \boldsymbol{c}$}}. \\
\end{array}
\]
\end{enumerate}

\section{Generalized ratio and completeness}
\label{gen-ratio}

Of course abandoning the Archimedes-Eudoxus postulate (or weakening it) we have to generalize the Eudoxus-Euclid definition of equality of ratios. Newton does not mention explicitly any necessary modifications, but his geometric representors of fluents suggest a natural direction.
One can consider two geometric quantities $\boldsymbol{a}$ and 
$\boldsymbol{b}$ as \emph{incomparable} if there is no geometric quantity measured by $\boldsymbol{a}$ and by $\boldsymbol{b}$. We call
$\boldsymbol{a}$ and by $\boldsymbol{b}$ \emph{comparable} if they
have common set of quantities measured by them, and moreover
if their decompositions into incomparable components are component-wise comparable: for any decomposition $\boldsymbol{a} = \sum \boldsymbol{a}_i$ into incomparable components $\boldsymbol{a}_i$ there exist (exactly one 
if $\boldsymbol{a}$ or $\boldsymbol{b}$ is an order unit) decomposition 
$\boldsymbol{b} = \sum \boldsymbol{b}_i$ into incomparable components 
$\boldsymbol{b}_i$ such that $\boldsymbol{a}_i$ measures or is measured by 
$\boldsymbol{b}_i$. 

Note that the following condition is trivially fulfilled if the Archimedes-Eudoxus postulate is valid (in the ordinary Eudoxus theory of ratios):
\emph{For any two geometric quantities $\boldsymbol{a}$ and $\boldsymbol{a'}$ -- and thus for any two $\boldsymbol{a}$ and $\boldsymbol{a'}$ which may compose a ratio as in this case all of them do  -- we have:}

\begin{enumerate} 
\item[]
\emph{for every fraction $k/l > 0$ there exist natural $m,n$such that}
\[
k/l \, \boldsymbol{a} \leq n\boldsymbol{a'} - m\boldsymbol{a} \leq k/l \, \boldsymbol{a}
\] 
\end{enumerate}
and actually the existence of a ratio $\boldsymbol{a':a}$ for every two quantities 
$\boldsymbol{a}$ and $\boldsymbol{a'}$
is nothing else but the celebrated \emph{completeness}
of geometric quantities. In more arithmetic terms and more in the spirit
of Dedekind it means that 
\begin{equation}\label{Dedekind}
\boldsymbol{a'} = \lambda \boldsymbol{a} \,\,\,\textrm{\emph{for}} \,\,\,
\lambda = \inf\{m/n: n\boldsymbol{a'} < m \boldsymbol{a}\} 
\end{equation}
so that the ratio $\boldsymbol{a':a}$ may be represented arithmetically
by the real number $\lambda$ (Dedekind's cut), where $\inf$ is with
respect to the natural order in the space of fractions, thus Dedekind's
construction of real number is purely arithmetic. But
keeping the geometric idea of Newton we rewrite (\ref{Dedekind}) in the 
following manner
\begin{equation}\label{Newton}
\boldsymbol{a'} = \inf_{n \boldsymbol{a'} < m \boldsymbol{a}} 
\{m/n \boldsymbol{a}\}, 
\end{equation}
where in this case $\inf$ is in the space of geometric objects, which
in Newton's parlance means that $\boldsymbol{a'}$ compose a ratio 
$\boldsymbol{a':a}$ with $\boldsymbol{a}$, which in \emph{Principia} was called
by him \emph{last ratio} among the ratios $m/n \boldsymbol{a:a}$
with $m,n$ verifying $n\boldsymbol{a'} < m\boldsymbol{a}$. Of course this example where Archimedes-Eudoxus principle is fulfilled may seem trivial, and not sufficiently suggestive, but we give
now the general definitions and then compare them with the corresponding passages from Newton's \emph{Principia}, which inspired them. 

Thus, inspired by Newton we assume 
\begin{enumerate}
\item[]
{\bf Generalized completeness} \, \emph{Two geometric 
quantities $\boldsymbol{a}$ and $\boldsymbol{a'}$ may compose
a ratio $\boldsymbol{a':a}$ if the following two conditions 
are fulfilled} 
\item[1)]
\emph{$\boldsymbol{a}$ is an order unit and if 
$\boldsymbol{a}$ and $\boldsymbol{a'}$ are \emph{comparable}: for any decomposition of $\boldsymbol{a'} = \boldsymbol{a'}_1 + \boldsymbol{a'}_2$ into incomparable components $\boldsymbol{a'}_i$ there exists unique decomposition $\boldsymbol{a} = \boldsymbol{a}_1 + \boldsymbol{a}_2$  into incomparable components $\boldsymbol{a}_i$, such that $\boldsymbol{a'}_i$ is measured by $\boldsymbol{a}_i$, $i \in \{1,2\}$.}
\item[2)] \emph{There exist a decreasingly directed net of decompositions
 such as in 1) (denote the set of those decompositions by $\mathcal{D}$) 
such that 
\[
\boldsymbol{a'} = \inf_{n_i \boldsymbol{a'}_i < m_i \boldsymbol{a}_i}
\{m_1/n_1 \boldsymbol{a}_1
+ \ldots + m_k/n_k \boldsymbol{a}_k \}
\]
where $\inf$ ranges over the decompositions $\boldsymbol{a}_1
+ \ldots + \boldsymbol{a}_k \in \mathcal{D}$}
\end{enumerate}
We say that the family (not necessary sequence) of decompositions is decreasingly directed if for any two decompositions of that family there exist a third such that
every component of the first two decompositions is a sum of incomparable components
of the third decomposition.

Of course it is important that for any other family of decreasingly directed  decompositions which have an infimum such as in condition 2) the infimum should be equal
to $\boldsymbol{a'}$. In subsequent Sections we give many interesting examples in 
which this actually holds and the definition makes sense. 

One can then compare two ratios $\boldsymbol{a':a}$ and $\boldsymbol{c':c}$ whenever for every such decompositions $\boldsymbol{a'}_i$ and $\boldsymbol{a}_i$ for the first ratio $\boldsymbol{a':a}$ there exist corresponding decompositions 
$\boldsymbol{c'}_i$ and $\boldsymbol{c}_i$ for the second such
that  $\boldsymbol{a}_i$ measures and is measured by $\boldsymbol{c}_i$,
then we put
\begin{enumerate}
\item[]
{\bf Generalized Eudoxus-Euclid definition} \, \emph{
$\boldsymbol{a':a}$ and $\boldsymbol{c':c}$ are said to be equal iff for any decomposition 
$\boldsymbol{a} = \boldsymbol{a}_1 + \boldsymbol{a}_2$ and the decompositions $\boldsymbol{c} = \boldsymbol{c}_1 + \boldsymbol{c}_2$,
$\boldsymbol{a'} = \boldsymbol{a'}_1 + \boldsymbol{a'}_2$,
$\boldsymbol{c'} = \boldsymbol{c'}_1 + \boldsymbol{c'}_2$ corresponding to it, and for arbitrary natural numbers $n_i, m_i$ ($i \in \{1,2\}$)}
\[
\begin{array}{l}
\textrm{(I)} \,\,\, \textrm{\emph{from $n_i \boldsymbol{a'}_i 
< m_i \boldsymbol{a}_i$ it follows $n_i \boldsymbol{c'}_i < m_i 
\boldsymbol{c}_i$}} \\
\textrm{\emph{and}} \\
\textrm{(II)} \,\,\, \textrm{\emph{from $n_i \boldsymbol{a'}_i \nless m_i \boldsymbol{a}_i$ it follows $n_i \boldsymbol{c'}_i \nless 
m_i \boldsymbol{c}_i$}}. \\

\end{array}
\]    
\end{enumerate}

Generalized completeness condition is inspired by the following Lemma II,
Book I of Newton's \emph{Principia}, which we quote in Motte's 
\cite{motte} translation:

\vspace*{0.5cm}

\begin{enumerate}
\item[]
''\emph{If in any figure AacE, terminated by the right lines Aa, AE, and the curve
acE, there be inscribed any number of parallelograms Ab, Bc, Cd, e.t.c.,
comprehended under equal bases AB, BC, CD, e.t.c., and the sides, Bb,Cc, Dd, e.t.c.,
parallel to one side Aa of the figure; and the parallelograms aKbl, bLcm, cMdn,
e.t.c. are completed. Then if the breadth of those parallelograms be supposed to be 
diminished, and their number to be augmented in infinitum; I say, that the ultimate
ratios which the inscribed figure AKbLcMdD, the circumscribed figure AalbmcndoE,
and curvilinear figure AabcdE, will have to one another, are ratios of equality.}''
\end{enumerate}
The generalized Eudoxus-Euclid definition of equality of ratios is in turn inspired
by the following Lemma IV, Book I of Newton's \emph{Principia}:

\vspace*{0.5cm}

\begin{enumerate}
\item[]
''\emph{If in two figures AacE, PprT, you inscribe (as before) two ranks of parallelograms, an equal number in each rank, and, when their breadths are 
diminished in infinitum, the ultimate ratios of the parallelograms in one 
figure to those in the other, each to each respectively, are the same; I say,
that those two figures AacE, PprT, are to one another in that same ratio.}''
\end{enumerate}
Note that here \emph{figures} are geometric objects which represent geometrically
Newton's \emph{fluents}, which is consequently practised all over the \emph{Principa}.
That the figures represent fluents has important consequences, because fluents cannot be reduced just to the area of the figures and subject to the equivalence relation
by partitions as in the Archimedes' writings on quadrature,
and reduced to a space with the Archimedes-Eudoxus postulate. Fluxions are 
algebraic objects with more structure, which in particular may be added
in a distinguished manner reflected by the distinguished \emph{abscissa} 
and \emph{ordinate}. The order structure used by Newton reflects that 
along the ordinate, as well as the addition.  
That the ordinary Eudoxus theory of ratios must be generalized is seen by a simple
example of two figures (representing Newton's fluxions)  
which cannot be compared, although we can see that the general lines of 
reasoning presented in \emph{Principia} work pretty good in this case:
namely we can consider representors of two fluxions such that whenever
the ordinate of the first is non-zero the ordinate of the other is zero.
That Newton had in mind that more natural structure of fluxions to be
in agreement with the intended order structure can be seen already in the cited lemmas
as he distinguishes the two lines: abscissa and ordinate. The same is seen in all
other places of \emph{Principia}. But the intended order is even more explicitly expressed in the above Lemma IV, although hidden among the conditions 
assuring the equality of ratios. Thus Newton intended to construct such a 
ratio $\boldsymbol{a':a}$ of two figures $\boldsymbol{a'}$ and $\boldsymbol{a}$ representing his fluents, which determines $\boldsymbol{a'}$ uniquely as 
a representor of a fluent, provided we have $\boldsymbol{a':a}$ and 
$\boldsymbol{a}$. Of course no real number or its equivalent can match this
demand. Functional analysis is necessary to develop Newton's ingenious idea.        

The circumstance that in more general situation
(i.e. outside the regime of Archimedes-Eudoxus postulate) 
not every two ratios are immediately comparable 
along the Eudoxus-Euclid-type definition, mentioned to above, enhances
the role of the fact that the ratio is an operation acting in the space
of geometric quantities. Namely, two ratios $\boldsymbol{a':a}$ and 
$\boldsymbol{b':b}$  may also be compared as operations acting in the space. In natural situation the space is linear
over the reals. The fact that the space of geometric quantities which can be measured by a fixed quantity $\boldsymbol{b}$ is additively closed and that the composition of two ratios is well defined whenever each of them is well defined as operation acting in the space is expressed by the following two postulates implicitly used
in Euclid's \emph{Elements} (compare the comment of Heath \cite{elements}):
\begin{enumerate}
\item[]
{\bf Additivity postulate} \,\, \emph{If $\boldsymbol{b}$ measures 
$\boldsymbol{c}$ and $\boldsymbol{b}$
measures $\boldsymbol{d}$, then $\boldsymbol{b}$ measures the sum 
$\boldsymbol{c}+ \boldsymbol{d}$.}

\item[]
{\bf Composition postulate} \,\, \emph{If $\boldsymbol{b}$ measures 
$\boldsymbol{c}$ and $\boldsymbol{c}$ measures $\boldsymbol{d}$, 
then $\boldsymbol{b}$ measures $\boldsymbol{d}$.}
\end{enumerate} 
That the ratio as an operation acting on geometric quantities is additive is expressed
by the following theorem 
\[
\textrm{\emph{if}}  \,\,\, \boldsymbol{a':a} = \boldsymbol{b':b} \,\,\,
\textrm{\emph{then}} \,\,\, \boldsymbol{a':a} = \boldsymbol{b':b}
= \boldsymbol{(a'+b'):(a+b)},
\]
expressed as Proposition 12, Book V of Euclid \emph{Elements}, and which is quoted by
Aristotle (\emph{Eth. Nic.} v. 7, 1131 b 14) in the shortened form ``the whole is to the whole what each part is to each part (respectively)''; Newton in his 
\emph{Principia} uses the term \emph{compositio} to designate this property of ratios.
  
Thus the space $S_{\boldsymbol{b}}$ of geometric quantities which can be measured by a fixed quantity $\boldsymbol{b}$ generates a linear space 
$S_{\boldsymbol{b}} - S_{\boldsymbol{b}}$ in which a representation of the algebra of ratios acts. These postulates are sufficiently general to be kept in the more general situation suggested by Newton, with the natural proviso in the composition postulate that
$\boldsymbol{c}$ and $\boldsymbol{b}$ compose a ratio, as well as that $\boldsymbol{d}$ and $\boldsymbol{c}$ do, and that both order units 
$\boldsymbol{b}$  and $\boldsymbol{c}$ are comparable. In general the ratio  $\boldsymbol{a:b}$ transforming $\boldsymbol{b}$ into $\boldsymbol{a}$ is uniquely determined by $\boldsymbol{b}$ and $\boldsymbol{a}$ and by the 
surrounding geometric axioms as a linear operator in the linear space generated by $S_{\boldsymbol{b}}$. Thus the Eudoxus-Euclid-type definition of equality of ratios will suffice in determining fully the set of equivalence classes of ratios. Although there is a woe: the multiplication structure given by ratio composition will not always agree with the composition of ratios as linear operators acting in the linear space generated by 
$S_{\boldsymbol{b}}$ (though both actions of a ratio $\boldsymbol{d:c}$ agree in the subspace of $S_{\boldsymbol{b}}$ of positive elements comparable with the order unit $\boldsymbol{c}$ defining the ratio 
$\boldsymbol{d:c}$). The situation when they do not agree may happen only if the linear operator multiplication structure of ratios is non-commutative, although this is not always the case, even for the non-commutative linear-operator composition of ratios.  Ratios always commute whenever are comparable, and thus compose commutative sub algebras of ratios of comparable quantities, which can be commonly measured. (Compare the Jordan-Von Neumann-Wigner characterisation of
observable algebras \cite{neumann}.)       

\vspace*{0.5cm}

{\bf Remark} \,\,\, Of course abandoning the Archimedes-Eudoxus postulate
potentially one have many different ways of generalizing the notion of ratio,
but the proposed generalization of completeness seems natural. 
The proposed one may be called a ``point-wise completeness'' characteristic of measure theory or for weak closedeness of von Neumann operator algebras.  Is this the only structure which can be successfully treated within 
the proportion theory? We don't think so. However it is a very subtle and difficult
task to fine tune the completeness and to strength it just as to pick out
the ratios corresponding to ratios of ``smooth'' geometric objects interesting
enough. Newton, nonetheless, suggested a way to proceed with this task with his 
proportions (ratios) and suggested to use the general method of ratios to his
differentials, \emph{moments} as he calls them, and to geometric 
representors of his fluents and fluxions. 
Having established the ``smooth geometric objects'' (very non-trivial task)
we may than impose ``uniform-type completeness'' much stronger than the 
``point like'' with respect to the norm 
\[
\parallel \boldsymbol{a} \parallel = \inf_{-m/n \boldsymbol{u} < \boldsymbol{a}
< m/n \boldsymbol{u}} \{m/n > 0\}, \,\,\, 
\textrm{\emph{where $\boldsymbol{u}$ is an order unit}}
\]
characteristic of topology or of uniform or norm completeness of C*-sub algebras of operators in Hilbert space. But before embarking on
 differential and topology structures we prefer first to give some 
measure-theoretical examples and show how they match with the theory of ratios.

\section{First elementary example}
\label{example1}

Let us give simple example. We assume that the geometric space induced
by $S_{\boldsymbol{b}}$ is complete in the sense that for any fixed element 
$\boldsymbol{a}$ of $S_{\boldsymbol{b}}$ the order completion of the elements of the form $m\boldsymbol{a}/n$ does not leads us out of the space 
$S_{\boldsymbol{b}}$ and compose a subspace fulfilling the Archimedes-Eudoxus postulate. In short we assume that $S_{\boldsymbol{b}}$ generates a real linear space, with the action induced by the action of fractions $m/n$ discussed above, in which $S_{\boldsymbol{b}}$ forms a cone of positive elements:
\begin{enumerate}
\item[1)]
if  $\boldsymbol{a} \in S_{\boldsymbol{b}}$, then $\boldsymbol{a} \neq 0$

\item[2)]
if $\boldsymbol{a} \in S_{\boldsymbol{b}}$ and  $\boldsymbol{c} \in 
S_{\boldsymbol{b}}$, then  $\boldsymbol{a} + \boldsymbol{c} \in 
S_{\boldsymbol{b}}$,
\item[3)] 
if $\mathbb{R} \ni \lambda > 0$ and  $\boldsymbol{a} \in S_{\boldsymbol{b}}$, then  $\lambda\boldsymbol{a} \in S_{\boldsymbol{b}}$. 
\end{enumerate} 
We assume further and only for simplicity that the linear space 
$H = S_{\boldsymbol{b}} - S_{\boldsymbol{b}}$ generated by 
$S_{\boldsymbol{b}}$ is finite dimensional over the reals. It is convenient to use the space $S_{\boldsymbol{a}}$ of quantities measured by 
$\boldsymbol{a}$ even for more general $\boldsymbol{a}$, which may not
be any order unit. When $H$ is finite dimensional using 
the \emph{minimal quantities} is convenient: we say 
that $\boldsymbol{a}$ is \emph{minimal} iff from 
$n\boldsymbol{a} > \boldsymbol{b}$ for some natural $n$ it follows that
$m\boldsymbol{b} > \boldsymbol{a}$ for some natural $m$. It is easily
seen that $\boldsymbol{a} \in S_{\boldsymbol{b}}$ is \emph{minimal} 
if and only if $\boldsymbol{a}$ is not a sum of two incomparable elements of $S_{\boldsymbol{b}}$ (i.e. sum of two incomparable non-zero positive elements). Thus in finite dimensional $H$ any element of $S_{\boldsymbol{b}}$ is a sum of incomparable minimal elements of $S_{\boldsymbol{b}}$. It is easily seen that the linear subspace $S_{\boldsymbol{a}} 
- S_{\boldsymbol{a}}$ is one dimensional if $\boldsymbol{a}$ is minimal: indeed minimality condition for 
$\boldsymbol{a}$ means that $S_{\boldsymbol{a}}$ fulfils the  
Archimedes-Eudoxus postulate, thus 
$S_{\boldsymbol{a}} - S_{\boldsymbol{a}}$ form a one dimensional space
over the reals (Dedekind). If $\boldsymbol{a}$ and $\boldsymbol{a'}$
are two incomparable minimal elements then 
$S_{\boldsymbol{a}} - S_{\boldsymbol{a}}$ and $S_{\boldsymbol{a'}} - S_{\boldsymbol{a'}}$ are linearly independent.   

If instead of the Archimedes-Eudoxus postulate we assume the 
\begin{enumerate}
\item[]
{\bf Riesz additivity postulate} \,\,\, $S_{\boldsymbol{a} +\boldsymbol{c}} = S_{\boldsymbol{a}} +S_{\boldsymbol{c}}$, 
\end{enumerate}
then there exist positive elements $\boldsymbol{a}_i$, $1 \leq i \leq \dim H$, forming a base in $S_{\boldsymbol{b}}$, unique up to proportionality: namely 
we take the order unit $\boldsymbol{b}$ and decompose it inductively
into minimal elements $\boldsymbol{b} = \boldsymbol{a}_1 + \dots + \boldsymbol{a}_n$ (the process must be possible for finite $n$ because of the assumed finite dimensionality of $H$). Then one observe that 
$S_{\boldsymbol{b}} = S_{\boldsymbol{a}_1 + \dots + \boldsymbol{a}_n}
=  S_{\boldsymbol{a}_1} + \dots + S_{\boldsymbol{a}_n}$, thus 
$\boldsymbol{a}_i, 1 \leq i \leq n$ composes a basis, so that $n = \dim H$. One can then introduce the Hilbert space structure into the linear space $H$ generated by $S_{\boldsymbol{b}}$, just assuming one of such canonical bases to be orthonormal. Then $S_{\boldsymbol{b}}$ becomes a self-dual cone of positive elements which is facially homogeneous in the sense of \cite{Connes} (see definition below). 

In this situation any two order units $\boldsymbol{a}$ and
$\boldsymbol{b}$ are comparable: indeed $\boldsymbol{b}$ is an
order unit if and only if its components in the canonical orthonormal
basis $\boldsymbol{a}_i$ are positive and non-zero and any two order units may be uniquely decomposed in the canonical basis $\boldsymbol{a}_i$ of (minimal) incomparable orthogonal elements. Their comparability then follows from definition, as any two elements of $S_{\boldsymbol{a}_i}$ with fixed $i$ are comparable as $\boldsymbol{a}_i$ is minimal and forms the space 
$S_{\boldsymbol{a}_i}$ fulfilling the Archimedes-Eudoxus postulate. 
Two ratios $\boldsymbol{b':b}$ and $\boldsymbol{a':a}$ are equal 
if and only if the first, transforming $\boldsymbol{b}$ 
into $\boldsymbol{b'}$, multiplies
the components of $\boldsymbol{b}$ in the canonical basis $\boldsymbol{a}_i$
by the same real numbers as the ratio $\boldsymbol{a':a}$ transforming
$\boldsymbol{a}$ into $\boldsymbol{a'}$. Thus the ratio $\boldsymbol{b':b}$ is well defined as a linear operator in $H$, uniquely determined by the condition that it transforms $\boldsymbol{b}$ into $\boldsymbol{b'}$, and every ratio acts as an linear operator which is diagonal in the canonical
basis $\boldsymbol{a}_i$. Thus in this case the algebra of ratios is a commutative algebra of operators in the Hilbert space $H$ with every order unit being a cyclic and separating vector for the algebra.

Recall that if we consider the complexification $H_{\mathbb{C}}= H \oplus iH = S_{\boldsymbol{b}} - S_{\boldsymbol{b}} + i S_{\boldsymbol{b}} -i S_{\boldsymbol{b}}$ of the real Hilbert space $H$ then the algebra
of ratios may be considered the real part of the commutative algebra of 
$\mathbb{C}$-linear operators, diagonal in the canonical basis 
$\boldsymbol{a}_i$, restricted to the real subspace $H$; 
although we may consider also the operators of the algebra as extended over the whole complexified space $H_{\mathbb{C}}$. Thus the algebra of ratios may be viewed as the self-adjoint part of the commutative (von Neumann)
algebra of linear operators in $H_{\mathbb{C}}$ diagonal in the orthonormal
basis $\boldsymbol{a}_i$. 

\emph{Self-duality} of the cone $S_{\boldsymbol{b}}$
means that for every $\boldsymbol{a} \in H$ there exists a unique decomposition called the Jordan decomposition of $\boldsymbol{a}$
such that $\boldsymbol{a} = \boldsymbol{a}_+ - \boldsymbol{a}_-$,
with $\boldsymbol{a}_+, \boldsymbol{a}_- \in S_{\boldsymbol{b}}$ and  
$\langle \boldsymbol{a}_+,\boldsymbol{a}_- \rangle = 0$. 
Here $\langle \cdot,\cdot \rangle$ denote the inner product in $H$. This
property is equivalent (independently if $H$ is or is not finite dimensional -- a consequence of the Hahn-Banach theorem) to the equality  
$S_{\boldsymbol{b}} = \{\boldsymbol{x} \in H: \langle \boldsymbol{x},\boldsymbol{a} \rangle \geq 0 \,\, \forall \boldsymbol{a} 
\in S_{\boldsymbol{b}} \}$ and it is this equality which is commonly called
\emph{self-duality} of $S_{\boldsymbol{b}}$. One immediately see that in this case $S_{\boldsymbol{b}}$ is self-dual in the sense of the last equality also as a positive cone in $H_{\mathbb{C}}$.    

A subset $F \in S_{\boldsymbol{b}}$ is called a \emph{face} of $S_{\boldsymbol{b}}$ if it is a cone and if from 
$0 < \boldsymbol{x} \leq \boldsymbol{a}$, and $\boldsymbol{a} \in F$ it follows that 
$\boldsymbol{x} \in F$. For every face $F$ there exists an orthogonal face (we use self-duality of the cone $S_{\boldsymbol{b}}$) 
$F^{\bot} = \{\boldsymbol{x} \in S_{\boldsymbol{b}}: \,\,
 \boldsymbol{x} \bot \boldsymbol{a} \,\, \textrm{for very} \,\, 
\boldsymbol{a} \in F \}$. A cone 
$S_{\boldsymbol{b}}$
is called \emph{facially homogeneous} if for any face $F$ the operator
$P_F - P_{F^{\bot}}$ is a derivation of $S_{\boldsymbol{b}}$, where
$P_F$ is the orthogonal projection operator on the linear subspace generated 
by the face $F$. For every positive quantity $\boldsymbol{a}$, i.e.
belonging to $S_{\boldsymbol{b}}$, we may consider a face 
$\langle \boldsymbol{a} \rangle$ generated by it:
intersection of all faces containing $\boldsymbol{a}$. 
Recall that as a result of the additivity postulate
of Sect.~\ref{gen-ratio} the face $\langle \boldsymbol{a} \rangle$ is nothing else but the space $S_{\boldsymbol{a}}$ of (positive) quantities 
measured by $\boldsymbol{a}$: 
$S_{\boldsymbol{a}} = \langle \boldsymbol{a} \rangle$.

We call a $\mathbb{C}$-linear (resp. $\mathbb{R}$-linear) operator $\delta$ acting in 
$H_{\mathbb{C}}$ (resp. in $H$) a \emph{derivation of the cone 
$S_{\boldsymbol{b}}$} whenever $e^{t\delta}$ transforms 
$S_{\boldsymbol{b}}$ into $S_{\boldsymbol{b}}$ for every real $t$ ($e^{t\delta}$ transforms the cone into the cone).

As follows from \cite{Connes} the complex commutative (von Neumann) 
algebra of $\mathbb{C}$-linear operators in $H_{\mathbb{C}}$, diagonal in the canonical basis $\boldsymbol{a}_i$ whose self-adjont part is nothing
else but our algebra of ratios, is equal to the set of derivations of
the cone $S_{\boldsymbol{b}}$  with the multiplication structure given by
the ordinary operator composition. 

\vspace*{0.5cm}

{\bf Remark} \,\,\, A well known method of Riesz \cite{riesz1} shows that Riesz additivity postulate is equivalent to the assumption that the linear space 
$H= S_{\boldsymbol{b}} - S_{\boldsymbol{b}}$  form a lattice with the order $<$: for any $\boldsymbol{c}, \boldsymbol{d} \in H$ there exist 
$\sup(\boldsymbol{c}, \boldsymbol{d})$ and 
$\inf(\boldsymbol{c},\boldsymbol{d})$. In this case one can put 
$\boldsymbol{x}_+ = \sup(\boldsymbol{x}, 0)$ and 
$\boldsymbol{x_-} = \sup(-\boldsymbol{x},0) = - \inf\boldsymbol{x},0)$
to get a Jordan decomposition of $\boldsymbol{x}$. In this lattice case were
there exist positive (minimal) elements mutually incomparable and orthogonal, which span the whole space $H$,
they give the minimal decomposition into a difference of two positive elements: for any decomposition  $\boldsymbol{x} = \boldsymbol{x'_+} 
-\boldsymbol{x'_-}$ with $\boldsymbol{x'_{\pm}} \in S_{\boldsymbol{b}}$
 we have $\boldsymbol{x'_+} \geq \boldsymbol{x_+}$ and 
$\boldsymbol{x'_-} \geq \boldsymbol{x_-}$. There does not exist such minimal
decomposition if there are no positive elements, i.e. elements
of $S_{\boldsymbol{b}}$,  mutually incomparable and orthogonal, which span $S_{\boldsymbol{b}}$ and thus the whole linear space $H$.

\section{Second elementary example}
\label{example2}

One can consider further generalizations of the previous (first) 
elementary example, using the results of \cite{Connes}, \cite{I}. 

Namely we interpret a general self-dual, facially homogeneous cone of positive elements in a real Hilbert space $H$ giving the relation $<$, keeping finite dimensionality of $H$, as $S_{\boldsymbol{b}}$. Two elements are by definition \emph{incomparable} in the sense of the order 
structure $<$ given by the cone of positive elements according to the definition of Sect.~\ref{gen-ratio} and, in addition, if they are orthogonal, as suggested by the first elementary example of Sect.~\ref{example1}. Because of finite dimensionality every element $\boldsymbol{a}$ of 
$S_{\boldsymbol{b}}$ may be decomposed into a finite sum of positive
incomparable and minimal elements $\boldsymbol{a}_i$, but in general 
if the Riesz additivity postulate is not valid there does not exist any 
base of minimal positive elements of $S_{\boldsymbol{b}}$, which span
$S_{\boldsymbol{b}}$ and $H= S_{\boldsymbol{b}} - S_{\boldsymbol{b}}$.
In this case the algebra of ratios is non commutative.
Moreover if the Riesz additivity is not fulfilled, 
minimality of $\boldsymbol{a}_i$ means that
 there do not exist
incomparable elements $\boldsymbol{a}_{i1}$ and $\boldsymbol{a}_{i2}$ such that
$\boldsymbol{a}_i = \boldsymbol{a}_{i1} + \boldsymbol{a}_{i2}$; thus for no decomposition $\boldsymbol{a}_i = \boldsymbol{a}_{i1} + \boldsymbol{a}_{i2}$ of
$\boldsymbol{a}_i$ can the elements 
$\boldsymbol{a}_{i1}$ and $\boldsymbol{a}_{i2}$ be separated by 
disjoint faces (having $0$ as the only common 
element).  This means, as we will see below, that the face $\langle \boldsymbol{a}_i \rangle$ generated by 
$\boldsymbol{a}_i$, i.e. 
the set $S_{\boldsymbol{a}_i}$ of elements measured by $\boldsymbol{a}_i$,
is minimal, which in turn means that there are no faces of the cone
$S_{\boldsymbol{b}}$ properly contained in $\langle \boldsymbol{a}_i \rangle$ which are orthogonal (or simply there are no faces properly contained in $\langle \boldsymbol{a}_i \rangle$ because to every face there exists the orthoginal one). Only if the Riesz additivity is fulfilled, i.e. when the order incomparability coincides with the orthogonality as in the preceding example,
one can immediately see that the minimal face 
$\langle \boldsymbol{a}_i \rangle$ generated by $\boldsymbol{a}_i$ is one dimensional.\footnote{Although there are many interesting examples of finite dimensional self dual cones, facially homogeneous, not fulfilling the Riesz additivity, which still have the property that minimal faces are one dimensional.}
We prefer here to stay on general grounds not assuming that the minimal
faces are one dimensional. 
But the ratios can still be  uniquely characterized by self adjoint derivations of the cone $S_{\boldsymbol{b}}$, just as in the previous elementary example of Sect.~\ref{example1}. In this non commutative
case there are two further possibilities:
\begin{enumerate}
\item[1)]
If in addition the cone is \emph{orientable} in the Connes sense of
\cite{Connes}, then the action of the ratios agrees with their action as linear operators, and the ratios compose the self-dual part of a von Neumann algebra of operators in $H_{\mathbb{C}}$ with multiplication given
by the ordinary operator composition. Although the von Neumann algebra is substantially smaller that the algebra of all derivations of 
$S_{\boldsymbol{b}}$, in contradistinction to the elementary commutative example.
\item[2)]
If the cone is not orientable then the composition of ratios gives
a structure of Jordan-Banach algebra, which does not agree with the 
action of ratios as linear operators in $H$.
\end{enumerate}

Before we proceed further we should explain the meaning of 
\emph{orientability} of the cone $S_{\boldsymbol{b}}$ in the sense of Alain Connes. One can see  that derivations of the cone compose a Lie algebra with the Lie structure given by the ordinary commutator. If the quotient of the Lie algebra by its (Lie) center has a (Lie) complex structure, then Connes calls the cone 
$S_{\boldsymbol{b}}$ \emph{orientable}.

The fact that ratios are uniquely characterized by self adjoint derivations of the cone $S_{\boldsymbol{b}}$, viewed as $\mathbb{R}$-linear operators  acting in $H$ may be expressed more precisely in the following two theorems:
\begin{twr}
For every ratio $\boldsymbol{a':a}$ there exist a unique self adjoint derivation $\delta$ of $S_{\boldsymbol{b}}$ in $H$ such that
$\delta \boldsymbol{a} = \boldsymbol{a'}$. For every ratio 
$\boldsymbol{c':c}$ such that $\boldsymbol{c':c} = \boldsymbol{a':a}$ we have: $\delta \boldsymbol{c} = \boldsymbol{c'}$. \label{twr1}
\end{twr}
We have also the converse
\begin{twr}
Every self adjoint derivation $\delta$ of $S_{\boldsymbol{b}}$ defines a ratio $\boldsymbol{a':a}$, where $\boldsymbol{a}$ is an order unit which may be decomposed into incomparable components lying in spectral faces of 
$\delta$ and $\boldsymbol{a'} = \delta \boldsymbol{a}$. Whenever a ratio 
$\boldsymbol{c':c}$ is comparable with $\boldsymbol{a':a}$ and
$\delta \boldsymbol{c} = \boldsymbol{c'}$, then we have
$\boldsymbol{c':c} = \boldsymbol{a':a}$. \label{twr2}
\end{twr}

The proof in this elementary finite dimensional
case requires only few elementary steps. 
It is convenient to introduce \emph{facial derivative} 
 $\delta_F = \frac{1}{2}(1 + P_F - P_{F^{\bot}})$ 
for every face $F$. Note that by assumption $S_{\boldsymbol{a}_1} = 
\langle \boldsymbol{a}_1 \rangle$
and $S_{\boldsymbol{a}_2} = \langle \boldsymbol{a}_1 \rangle$ 
are disjoint (with $0$ as the only common element) faces if $\boldsymbol{a}_1$ and 
$\boldsymbol{a}_1$ are incomparable. Because $\boldsymbol{a}_2 \in 
{\langle \boldsymbol{a}_1 \rangle}^{\bot}$ and $\langle \boldsymbol{a}_2 \rangle$ is the smallest face containing $\boldsymbol{a}_2$, then
$\langle \boldsymbol{a}_2 \rangle \subseteq 
{\langle \boldsymbol{a}_1 \rangle}^{\bot}$; thus $\langle \boldsymbol{a}_1 \rangle \bot \langle \boldsymbol{a}_2 \rangle$ for incomparable elements
$\boldsymbol{a}_1$ and $\boldsymbol{a}_2$.  Let $\boldsymbol{a'}_i$
and $\boldsymbol{a}_i$, $1 \leq i \leq n$, be the respective 
decompositions of $\boldsymbol{a'}$ and $\boldsymbol{a}$ into 
incomparable components. We can assume
them of course to be minimal components in this finite dimensional case. 
For every such decomposition we define the derivation
\begin{equation}\label{deriv1}
\delta_{\boldsymbol{a}_1 + \dots + \boldsymbol{a}_n} = 
\lambda_1 \delta_{\langle \boldsymbol{a}_1 \rangle}
+ \ldots + \lambda_n \delta_{\langle \boldsymbol{a}_n \rangle},
\end{equation}   
where
\begin{equation}\label{lambda}
 \lambda_i = \inf \{m_i/n_i : n_i \boldsymbol{a'}_i < m_i \boldsymbol{a}_i ,
\,\,\, m_i , n_i \in \mathbb{N}\},
\end{equation}
and $\langle \boldsymbol{a}_i \rangle , \langle \boldsymbol{a'}_i \rangle$
are the respective faces generated by $\boldsymbol{a}_i$ and 
$\boldsymbol{a'}_i$. Because of the assumed generalized completeness
(Sect. ~\ref{gen-ratio}), minimality of 
$\langle \boldsymbol{a}_i \rangle$ and 
of closedeness of the Hilbert space $H$, we have
$\lambda_i \boldsymbol{a}_i = \boldsymbol{a'}_i$. Because
$\delta_{\boldsymbol{a}_1 + \dots + \boldsymbol{a}_n} \boldsymbol{a} = 
\lambda_1 \boldsymbol{a}_1 + \dots + \lambda_n \boldsymbol{a}_n$,
then $\delta_{\boldsymbol{a}_1 + \dots + \boldsymbol{a}_n} \boldsymbol{a}= 
\boldsymbol{a'}$. For any other such decompositions (if possible) we 
obtain the corresponding derivations, which maps $\boldsymbol{a}$ into 
$\boldsymbol{a'}$, but any two such derivations must be equal
as the unit order $\boldsymbol{a}$ is separating (and cyclic)
for the set of derivations, see \cite{I}, Corollary II.5. Thus $\delta$
is uniquely determined by the ratio $\boldsymbol{a':a}$, if it is well
defined at all. That the derivation $\delta$ is well defined by the ratio
$\boldsymbol{a':a}$ means that any two elements $\boldsymbol{c'}$ and $\boldsymbol{c}$ composing a ratio $\boldsymbol{c':c}$
equal to $\boldsymbol{a':a}$ define the same derivation. But that the derivation is well defined follows from the generalized Eudoxus-Euclid
definition of equality of ratios (see the Sect ~\ref{intro}). Indeed,
for the corresponding decompositions $\boldsymbol{c'}_i$ and 
$\boldsymbol{c}_i$ of $\boldsymbol{c'}$ and $\boldsymbol{c}$ the existence of which is assured by that definition, we have $\langle \boldsymbol{c}_i \rangle = \langle \boldsymbol{a}_i \rangle$ with exactly the same
corresponding $\lambda_i$, thus $\delta_{\boldsymbol{a}_1 + \dots + \boldsymbol{a}_n}$ is equal to the corresponding 
$\delta_{\boldsymbol{c}_1 + \dots + \boldsymbol{c}_n}$ and $\delta$
is well defined. This ends the proof of Theorem ~\ref{twr1}. 

In the proof of Theorem ~\ref{twr2} we will use the following and easy 
(\cite{BI1}, lemma III.3)
\begin{lem}
If $\nu_1$ and $\nu_2$ are finite positive Borel measures on $\mathbb{R}$ such
that
\[
\int e^{t\lambda} \,d\nu_{1}(\lambda) 
\leq \int e^{t\lambda} \,d\nu_{2}(\lambda), \,\,\,
 \textrm{\emph{for all real $t$}},
\]
and if $\nu_2$ is concentrated on an (not necessarily closed) interval, then
$\nu_1$ is also concentrated on the same interval. \label{lem}
\end{lem}
Theorem ~\ref{twr2} follows from an equivalent of the ordinary spectral
theorem for derivations, compare \cite{BI1}. 
Let $P_{\lambda_1}, \ldots P_{\lambda_m}$ be the spectral projections
of $\delta$ corresponding to spectral values $\lambda_1, \ldots \lambda_m$.
We observe first that the set $F_{\lambda_i} = P_{\lambda_i}H \cap 
S_{\boldsymbol{b}}$ is a face.  Indeed, let $\pi(\lambda)$
be the increasing family of projectors giving the projection valued spectral measure
$d\pi(\lambda)$ of $\delta$ (of course purely discrete in this
finite dimensional case). Then $\boldsymbol{\xi} \in F_{\lambda_i}$
means that the spectral measure $ d\nu_{\boldsymbol{\xi}}(\lambda) 
= d\langle \boldsymbol{\xi}, \pi(\lambda)\boldsymbol{\xi} \rangle$ 
is concentrated at the point $\lambda_i$.
Let $\boldsymbol{\eta}$ be such that $0 \leq \boldsymbol{\eta} \leq \boldsymbol{\xi}
\in F_{\lambda_i}$. Then $0 \leq \langle e^{t\delta}\boldsymbol{\eta},\boldsymbol{\eta}
\rangle \leq \langle e^{t\delta} \boldsymbol{\xi}, \boldsymbol{\xi} \rangle$,
for all real $t$. Thus by spectral theorem 
\[
\int e^{t\lambda} \,d\nu_{\boldsymbol{\eta}}(\lambda) 
\leq \int e^{t\lambda} \,d\nu_{\boldsymbol{\xi}}(\lambda), \,\,\,
 \textrm{\emph{for all real $t$}},
\] 
which is possible only if the measure $d\nu_{\boldsymbol{\eta}}$ is also
concentrated at the point $\lambda_i$, therefore 
$\boldsymbol{\eta} \in F_{\lambda_i}$ and $F_{\lambda_i}$ is a face.
Let $\boldsymbol{a}_i$ be its order unit: $\langle \boldsymbol{a}_i \rangle
= F_{\lambda_i}$. Observe that $F_{\lambda_i}$ cannot be $\{0\}$ for all
$\lambda_i$, as for any subspace $L$ of $H$ either $L\cap S_{\boldsymbol{b}}
\neq \{0\}$ or $L^\bot \cap S_{\boldsymbol{b}}
\neq \{0\}$ (lemma III.4 of \cite{BI1}). Note that $\boldsymbol{a} =
\boldsymbol{a}_1 \ldots + \boldsymbol{a}_n$ is an order unit if 
order units $\boldsymbol{a}_i$ of all $F_{\lambda_i} \neq \{0\}$ are here
included. Indeed by construction
$\langle\boldsymbol{a}\rangle^\bot = \{0\}$, and thus $\boldsymbol{a}$ must be
an order unit (in particular for finite dimensional cone weak interior points of the cone and order units coincide, see \cite{I}, as they have non-empty topological interior). Therefore $\delta \boldsymbol{a} 
= \big( \lambda_1 \delta_{\langle \boldsymbol{a}_1 \rangle} + \ldots 
+ \lambda_n \delta_{\langle \boldsymbol{a}_n \rangle} \big) \boldsymbol{a}$, and because 
$\boldsymbol{a}$ is separating for derivations we have   
\[
\delta = \lambda_1 \delta_{\langle \boldsymbol{a}_1 \rangle} + \ldots 
+ \lambda_n \delta_{\langle \boldsymbol{a}_n \rangle}. 
\] 
Let $\delta \boldsymbol{a} = \boldsymbol{a'} = \lambda_1 \boldsymbol{a}_1 + \ldots 
+ \lambda_n \boldsymbol{a}_n$.
We define $\boldsymbol{a':a}$ as the ratio corresponding to $\delta$.
Recall that by the uniqueness of the spectral family $\pi(\lambda)$ (equivalently
the projectors $P_{\lambda_i}$) any other ratio $\boldsymbol{c':c}$ defined
by $\delta$ is comparable with $\boldsymbol{a':a}$ as it is constructed by choosing
other order units $\boldsymbol{c}_i$ of $F_{\lambda_i}$. But then 
$\langle\boldsymbol{c}_i \rangle = \langle \boldsymbol{a}_i \rangle$ and $\delta$
act as a multiplication by $\lambda_i$ within $\langle\boldsymbol{c}_i \rangle = \langle \boldsymbol{a}_i \rangle$. Thus the ratios $\boldsymbol{a':a}$ 
and $\boldsymbol{c':c}$ are equal in the sense of the generalized definition of 
equality of ratios. Theorem ~\ref{twr2} thus follows.

Having proved the Theorems ~\ref{twr1} and ~\ref{twr2}, the  two statements 
1) and 2) at the beginning of this Sect. are now immediate consequences
of the results presented in  \cite{I}, VI.2, restricted to finite 
dimension. 

\vspace*{0.5cm}

{\bf Remark} \,\,\, Consider the Hilbert subspace $H_{\boldsymbol{a':a}}
= H_{\langle \boldsymbol{a}_1 \rangle} \oplus \ldots \oplus H_{\langle \boldsymbol{a}_n \rangle}$ generated by the faces $\langle \boldsymbol{a}_i \rangle$ (i.e. each direct summand $H_{\langle \boldsymbol{a}_i \rangle} = \langle \boldsymbol{a}_i \rangle
-\langle \boldsymbol{a}_i \rangle$ is the Hilbert subspace generated by the face 
$\langle \boldsymbol{a}_i \rangle$). Note that not every vector 
$\boldsymbol{\eta}$ of the subspace $H_{\boldsymbol{a':a}}
= H_{\langle \boldsymbol{a}_1 \rangle} \oplus \ldots \oplus H_{\langle \boldsymbol{a}_n \rangle}$ may compose a ratio $\boldsymbol{\eta : a}$ with 
$\boldsymbol{a}$ when some among the minimal faces 
$\langle \boldsymbol{a}_i \rangle$ are not one dimensional but have higher 
dimension. Indeed only those $\boldsymbol{\eta} \in H_{\boldsymbol{a':a}}$ 
compose a ratio
with $\boldsymbol{a}$ and fulfil the condition 2) of the generalized completeness (Sect. ~\ref{gen-ratio}) which are obtainable from 
$\boldsymbol{a}$ by application of an operator which is a multiplication
operator within each direct summand subspace $H_{\langle \boldsymbol{a}_i \rangle}$. It is because $\boldsymbol{a'}$ is the image $\boldsymbol{a'} = \delta \boldsymbol{a}$ of 
$\boldsymbol{a}$ under the action of the operator
$\delta$ which act as step-wise diagonal operator, which acts as multiplication operator in each orthogonal subspace $H_{\langle \boldsymbol{a_i} \rangle}$ that the elements $\boldsymbol{a'}$ and $\boldsymbol{a}$
fulfil also the condition 2) of generalized completeness and do compose
a ratio $\boldsymbol{a':a}$.

\subsection{Some remarks with a view toward infinite dimension}
\label{hints}

It will be instructive to give the decisive steps of the above simple proof a form capable of further generalizations to infinite dimension of the next Sect. 
~\ref{example3}. First note that for $\boldsymbol{a}$ being an order unit and for its decomposition
$\boldsymbol{a} = \boldsymbol{a}_1 + \boldsymbol{a}_2$ into incomparable 
components $\boldsymbol{a}_1$ and $\boldsymbol{a}_2$ we easily check by an explicit inspection that
\[
\Big( \delta_{\langle \boldsymbol{a} \rangle} 
- \delta_{\langle \boldsymbol{a}_1 \rangle} \Big) \boldsymbol{a} = \boldsymbol{a}_2 .  
\]
Because $\delta_{\langle \boldsymbol{a}_2 \rangle} \boldsymbol{a} 
= \boldsymbol{a}_2$
we have 
\[
\Big( \delta_{\langle \boldsymbol{a} \rangle} 
- \delta_{\langle \boldsymbol{a}_1 \rangle} \Big) \boldsymbol{a} 
= \delta_{\langle \boldsymbol{a}_2 \rangle} \boldsymbol{a}.  
\]

Thus it follows that 
\[
\delta_{\langle \boldsymbol{a}_1 + \boldsymbol{a}_1 \rangle} \boldsymbol{a} 
= \Big( \delta_{\langle \boldsymbol{a}_1 \rangle} 
+ \delta_{\langle \boldsymbol{a}_2 \rangle} \Big) \boldsymbol{a}, 
\]
and because $\boldsymbol{a}$ is separating for derivations, we have
\begin{equation}\label{delta}
\delta_{\langle \boldsymbol{a}_1 + \boldsymbol{a}_1 \rangle} 
= \delta_{\langle \boldsymbol{a}_1 \rangle} 
+ \delta_{\langle \boldsymbol{a}_2 \rangle}, 
\end{equation}
whenever $\boldsymbol{a}_1 + \boldsymbol{a}_2$ is a decomposition
of an order unit $\boldsymbol{a}$ into incomparable components 
$\boldsymbol{a}_1$ and $\boldsymbol{a}_2$. Second, using the property
(\ref{delta}) we can rewrite the derivation (\ref{deriv1})
in the following form (assume in addition that the $\lambda_i$-s are
increasingly ordered, if not -- rearrange them):
\begin{multline}\label{deriv-sp}
\delta_{\boldsymbol{a}_1 + \dots + \boldsymbol{a}_n} 
= \lambda_1 \delta_{\langle \boldsymbol{a}_1 \rangle}
+ \ldots + \lambda_n \delta_{\langle \boldsymbol{a}_n \rangle} \\
= \lambda_1 \delta_{F(\lambda_1)} + \lambda_2 \Big( \delta_{F(\lambda_2)}
- \delta_{F(\lambda_1)}\Big) + \lambda_3 \Big(\delta_{F(\lambda_3)}
- \delta_{F(\lambda_2)}\Big) \\
\ldots + \lambda_n \Big(\delta_{F(\lambda_n)}
- \delta_{F(\lambda_{n-1})}\Big),
\end{multline}
where $\lambda_i \mapsto F(\lambda_i)$ is an increasing family of faces
$F(\lambda_i) = \langle \boldsymbol{a}_1 
+ \ldots + \boldsymbol{a}_i \rangle$. To them corresponds of course an increasing family of projectors $\lambda_i \mapsto P(\lambda_i) 
= P_{F(\lambda_i)}$,
called the spectral family of the derivation 
$\delta_{\boldsymbol{a}_1 + \dots + \boldsymbol{a}_n}$ as well as the spectral sum operator 
\begin{multline}\label{sp-op}
\lambda_1 P(\lambda_1) + \lambda_2 \Big( P(\lambda_2) - P(\lambda_1)\Big)
+ \lambda_3 \big( P(\lambda_3) - P(\lambda_2) \Big) \\
\ldots + \lambda_n \Big( P(\lambda_n) - P(\lambda_{n-1}) \Big).  
\end{multline}
Third, consider the ratio $\boldsymbol{a':a}$ of the Theorem
~\ref{twr1} and any decomposition $\boldsymbol{a} = \boldsymbol{a}_1
+  \ldots + \boldsymbol{a}_k$  of $\boldsymbol{a}$ into incomparable components $\boldsymbol{a}_i$ (not necessary minimal). 
Then take one of the components, say $\boldsymbol{a}_i$ and decompose it further $\boldsymbol{a}_i = \boldsymbol{a}_{i1} + \boldsymbol{a}_{i2}$
and consider the corresponding decomposition $\boldsymbol{a'}_i 
= \boldsymbol{a'}_{i1} + \boldsymbol{a'}_{i2}$ of $\boldsymbol{a'}_i $
and the corresponding 
\[
\begin{array}{l}
\lambda_i = \inf \{m_i/n_i : n_i \boldsymbol{a'}_i < m_i \boldsymbol{a}_i , \,\,\, m_i , n_i \in \mathbb{N}\},\\
\lambda_{i1} = \inf \{m_{i1}/n_{i1} : n_{i1} \boldsymbol{a'}_{i1} < m_{i1} 
\boldsymbol{a}_{i1} ,\,\,\, m_{i1} , n_{i1} \in \mathbb{N}\}, \\
\lambda_{i2} = \inf \{m_{i2}/n_{i2} : n_{i2} \boldsymbol{a'}_{i2} < m_{i2} 
\boldsymbol{a}_{i2} , \,\,\, m_{i2} , n_{i2} \in \mathbb{N}\},
\end{array}
\]
then
\begin{equation}\label{lambda-ineq}
\lambda_{i1} \leq \lambda_i \,\,\, \textrm{and} \,\,\, \lambda_{i2} \leq \lambda_i. 
\end{equation}
Consider now any decreasingly directed family (sequence) of decompositions, which means that in a decomposition of the family every component is further decomposed into sub-components of the subsequent decomposition. (Of course any such decreasingly directed sequence of decompositions ends up with a decomposition into minimal elements in this finite dimensional case.) Then the spectral sum operators (\ref{sp-op}) of the derivations  (\ref{deriv-sp}) corresponding to the decompositions are also 
decreasingly directed
in the ordinary order of selfadjoint operators\footnote{$A \leq B$
iff $B-A$ has positive spectrum.}, as follows from (\ref{lambda-ineq}). However the limiting spectral sum operator corresponding to decomposition into minimal components \emph{is not equal to the corresponding derivation} 
(\ref{deriv1}). Note also (\cite{BI1}, prop. III.1) that by lemma ~\ref{lem} if 
$\lambda \mapsto \pi(\lambda)$ is the increasing spectral
family $\lambda \mapsto \pi(\lambda)$ of projections of the derivation  operator (\ref{deriv1}) the sets
\begin{equation}\label{face}
 F(\lambda_i) = \{\boldsymbol{x} \in S_{\boldsymbol{b}}: 
\pi(\lambda_i)\boldsymbol{x} = \boldsymbol{x}\}
\end{equation}
are faces of the cone $S_{\boldsymbol{b}}$. (In general for 
some subsequent $\lambda_i$ and $\lambda_{i+1}$ it may happen that
the faces $F(\lambda_i)$ and $F(\lambda_{i+1})$ coincide.)    

However it is clear in this finite dimensional case that the spectral
sum operator (\ref{sp-op}) also determinates the corresponding derivation
(\ref{deriv-sp}) uniquely: for any self adjoint operator whose spectral
family has the property that the set (\ref{face}) is a face of the cone
$S_{\boldsymbol{b}}$ for every spectral value $\lambda_i$ determinates a derivation of $S_{\boldsymbol{b}}$ via the formula (\ref{deriv-sp}). Recall that the  spectral sum operator and the corresponding derivation $\delta_{\boldsymbol{a}_1 + \dots + \boldsymbol{a}_n} $ coincide in action on the order unit
$\boldsymbol{a} = \boldsymbol{a}_1 + \dots + \boldsymbol{a}_n$. An analogue construction of derivation out of the spectral sum operator is possible for infinite dimensional $H$. Namely, having an operator whose spectral family 
$\pi(\lambda)$ has the property that for every spectral value $\lambda$ the set $ F(\lambda) = \{\boldsymbol{x} \in S_{\boldsymbol{b}}: \pi(\lambda)\boldsymbol{x} = \boldsymbol{x}\}$ is a face of 
$S_{\boldsymbol{b}}$ one can construct a derivation of 
$S_{\boldsymbol{b}}$ as in the spectral theorem \cite{BI1}, using the spectral faces $F(\lambda)$ as in the analogue of the formula 
(\ref{deriv-sp}). This is important because in the infinite dimensional case it is the decreasingly directed family of spectral sum (integral) operators analogue to (\ref{sp-op}) and their limit which are more easily accessible than the corresponding derivations.

\section{Non-elementary example}
\label{example3}
Using the results of \cite{Connes}, \cite{BI1} and \cite{BI2}, 
summarized in \cite{I}, we can give a non-elementary example.

Namely, just consider $S_{\boldsymbol{b}}$ as a general 
self-dual, facially homogeneous cone as a cone of positive elements 
in a real Hilbert space $H$ giving the relation $<$, with $H$ not necessary finite dimensional (although separable, and we assume the topological interior of 
$S_{\boldsymbol{b}}$ to be non-empty). Two elements are 
\emph{incomparable} in the sense of the order structure $<$ given by the 
cone of positive elements according to the definition of Sect.~\ref{gen-ratio} and, in addition, if they are orthogonal, as suggested by the first elementary example of Sect.~\ref{example1}, but with some additional proviso, following from infinite dimensionality. Namely, in this infinite dimensional situation, it may happen that although there is no element which may be commonly measured by two orthogonal elements 
$\boldsymbol{a}_1$ and $\boldsymbol{a}_2$, it may nonetheless exits such $\boldsymbol{x}$ which can be approximated by the elements measured by $\boldsymbol{a}_1$ as well as the same $\boldsymbol{x}$ may be approximated by elements measured by $\boldsymbol{a}_2$. In this case (impossible for finite dimensional cone $S_{\boldsymbol{b}}$) elements 
$\boldsymbol{a}_1$ and $\boldsymbol{a}_2$ cannot be regarded as truly \emph{incomparable}. We therefore must define two elements $\boldsymbol{a}_1$ and $\boldsymbol{a}_2$ as \emph{incomparable} if and only if they are orthogonal and if there exist closed faces $F_1$ and $F_2$ of $S_{\boldsymbol{b}}$ such that 
$\boldsymbol{a}_i \in F_i $ and $F_1 \cap F_2 = \{0\}$.  Thus to the 
set $S_{\boldsymbol{a}}$ of all elements
measured by $\boldsymbol{a}$ it is natural to add all which may be approximated by them
in the norm of $H$.

In this situation the ratios can still be  uniquely 
characterized by self adjoint derivations of the cone 
$S_{\boldsymbol{b}}$, just as in the previous elementary examples of Sect.~\ref{example1} and ~\ref{example2}. If the Riesz additivity is not fulfilled, then the algebra of ratios will be non commutative, just as in the previous elementary example 
of Sect. ~\ref{example2}. There are two further possibilities in this case, exactly as in the second elementary example of Sect. ~\ref{example2}: 
\begin{enumerate}
\item[1)]
If in addition the cone is \emph{orientable} in the Connes sense of
\cite{Connes}, then the action of the ratios agrees with their action as linear operators, and the ratios compose the self-dual part of a von Neumann algebra of operators in $H_{\mathbb{C}}$ with multiplication given
by the ordinary operator composition. Although the von Neumann algebra is substantially smaller that the algebra of all derivations of 
$S_{\boldsymbol{b}}$, just as in the second elementary example of
Sect. ~\ref{example1}, and  contrary to the first elementary commutative example of Sect. ~\ref{example1}.
\item[2)]
If the cone is not orientable then the composition of ratios gives
a structure of Jordan-Banach algebra, which does not agree with the 
action of ratios as linear operators in $H$, just as for the finite dimensional case of the second elementary example.
\end{enumerate}

We assume the cone $S_{\boldsymbol{b}}$ to have non empty topological
interior in order to exclude some infinite dimensional pathology. 
Under this assumption the property of $\boldsymbol{u}$ to be an order unit, 
weak order unit (smallest face generated by $\boldsymbol{u}$ is equal to 
$S_{\boldsymbol{b}}$) or quasi-interior point of $S_{\boldsymbol{b}}$ ($\langle \boldsymbol{u} \rangle^\bot = 0$) coincide, see \cite{I}, proposition I.1.15, and all order units are separating for
selfadjoint derivations of $S_{\boldsymbol{b}}$, compare \cite{I}, corollary II.1.5.
Thanks to the separability assumption we have a sufficient supply of 
weak order units, and thus order units, as the set of weak order units is 
dense in $S_{\boldsymbol{b}}$, see e.g. \cite{BI2}, proposition 1.5, or \cite{I},
proposition I.1.16 i); and thus the set of order units is dense in 
$S_{\boldsymbol{b}}$, because the topological interior of the cone
$S_{\boldsymbol{b}}$ is non-empty. Besides the separability of $H$
allows us using the classical von Neumann theory of decompositions
in the proof of Theorems ~\ref{twr1} and ~\ref{twr2} for the
infinite dimensional example of this Section.

For any element $\boldsymbol{a} \in S_{\boldsymbol{b}}$ we can consider the 
intersection $F_{\boldsymbol{a}}$  of the faces which are \emph{closed} and contain 
$\boldsymbol{a}$, i.e. the smallest among the faces which are \emph{closed} and contain 
$\boldsymbol{a}$. In finite dimensional case this smallest face coincides with
$\langle \boldsymbol{a} \rangle$, i.e.the smallest face containing $\boldsymbol{a}$.
Note that for any \emph{incomparable} elements $\boldsymbol{a}_1$ and 
$\boldsymbol{a}_2$ the closed faces generated by them $F_{\boldsymbol{a}_1}$ and
$F_{\boldsymbol{a}_2}$ are orthogonal. Indeed $F_{\boldsymbol{a}_1} \cap F_{\boldsymbol{a}_1} = \{0\}$, and $\boldsymbol{a}_2 \in F_{\boldsymbol{a}_1}$, thus $F_{\boldsymbol{a}_2}\subseteq F_{\boldsymbol{a}_1}^\bot$, so that $F_{\boldsymbol{a}_1} \bot 
F_{\boldsymbol{a}_2}$. 

The algebraically generated face $\langle \boldsymbol{a} \rangle$, using $+, <$ and multiplication by a positive real is useless here because in general the topological closure of a face may not be a face. Because the closed face $F_{\boldsymbol{a}}$ generated by $\boldsymbol{a}$ is the immediate conceptual analogue of the 
``algebraically generated'' face $\langle \boldsymbol{a} \rangle$
(smallest face containing $\boldsymbol{a}$) of the previous Sect.
and we will not use algebraically generated faces here, then we denote the
closed face $F_{\boldsymbol{a}}$ generated by $\boldsymbol{a}$ by the same symbol
$\langle \boldsymbol{a} \rangle$. This notation will reflect the conceptual linkage of this Section with the previous one. Summing up
\begin{enumerate}
\item[]
\emph{In this Section 
\[
S_{\boldsymbol{a}} = \langle \boldsymbol{a} \rangle
\] 
means the smallest and closed face containing $\boldsymbol{a}$.}
\end{enumerate}
But recall that in the cited literature \cite{BI1}, \cite{BI2} the symbol
 $\langle \boldsymbol{a} \rangle$ denotes the smallest face containing 
$\boldsymbol{a}$ not necessary closed, and the notion of \emph{closed} face generated by a set of elements is not used there.

Thus an element $\boldsymbol{a}$ is minimal if and only if there does not exist two
incomparable elements $\boldsymbol{a}_1$ and $\boldsymbol{a}_2$ such that that
$\boldsymbol{a} = \boldsymbol{a}_1 + \boldsymbol{a}_2$; thus for no decomposition 
$\boldsymbol{a} = \boldsymbol{a}_1 + \boldsymbol{a}_2$ can the elements 
$\boldsymbol{a}_i$ be separated by orthogonal and closed faces. 

In this situation for every closed face $F = \bar{F}$ there exist an element
$\boldsymbol{x}$ such that $F = \langle \boldsymbol{x} \rangle = F_{\boldsymbol{x}}$.
Indeed, $F$ is a self dual cone in the Hilbert subspace
$H_F = F - F$ (\cite{I}, Lemma I.1.13 i)).
Then because $H$ is separable so is $H_F$, and there exist (even a set of such which is dense in $H_F$) a weak order unit $\boldsymbol{x}$ of $F$, so that $F = \langle \boldsymbol{x} \rangle$.      

For a ratio $\boldsymbol{a':a}$ consider a decreasingly directed
(in the sense of Subsect. ~\ref{hints}) family (sequence) $\mathcal{D}$
of decompositions of $\boldsymbol{a'}$ and corresponding to them
decompositions of $\boldsymbol{a}$. For every decomposition
$\boldsymbol{a} = \boldsymbol{a}_1 + \ldots + \boldsymbol{a}_n$
(and $\boldsymbol{a'} = \boldsymbol{a'}_1 + \ldots + \boldsymbol{a'}_n$) of that family consider the corresponding derivation (\ref{deriv-sp})
with $\lambda_i$ given by (\ref{lambda}) and the corresponding spectral
sum operator (\ref{sp-op}). Note that the derivation (\ref{deriv-sp})
and the corresponding operator (\ref{sp-op}) coincide in action on
$\boldsymbol{a}$ giving $\lambda_1 \boldsymbol{a}_1 + \ldots 
+ \lambda_n \boldsymbol{a}_n$. Because for any sub decomposition
we have the inequality (\ref{lambda-ineq}), then the spectral
sum operators (\ref{sp-op}) compose a decreasingly directed family
of operators (with the ordinary order: operator $A$ is greater then $B$ 
if $A-B$ has positive spectrum). Thus the family of spectral sum operators
(\ref{sp-op}) corresponding to the decreasingly directed family
of decompositions has a weak operator limit, say $Q$.
Similarly the the operators 
\[
\begin{array}{l}
-\lambda_1 P'(\lambda_1) + \lambda_2 \Big( P'(\lambda_1) - P'(\lambda_2)\Big)
+ \lambda_3 \big( P'(\lambda_2) - P'(\lambda_3) \Big) \\
\ldots + \lambda_n \Big( P'(\lambda_n-1) - P'(\lambda_{n}) \Big), 
\end{array}
\]
where $P'(\lambda_i) = P_{F(\lambda_i)^{\bot}}$ converge weakly, and thus
the corresponding derivations (\ref{deriv-sp}) converge weakly
to a selfadjoint operator $\delta$. Because
the set of self adjoint derivations is weakly closed then 
the limit $\delta$ must be a selfadjoit derivation. Moreover, strong operator closure and weak operator closure give rise to the same effect here thus, if $\delta\boldsymbol{a} = Q\boldsymbol{a} 
= \boldsymbol{a''}$, then $\lambda_1 \boldsymbol{a}_1 + \ldots 
+ \lambda_n \boldsymbol{a}_n$ converge in norm to the element 
$\boldsymbol{a''}$.

Next, for every decomposition $\boldsymbol{a} = \boldsymbol{a}_1 + \ldots + \boldsymbol{a}_n$ of our decreasingly directed family (sequence) of decompositions   consider the Hilbert space 
\begin{equation}\label{H-dec}
H_{\langle \boldsymbol{a}_1 \rangle} \oplus \ldots \oplus H_{\langle \boldsymbol{a}_n \rangle}. 
\end{equation}
In this way we get a decreasingly directed net of Hilbert spaces
\begin{multline}\label{H-net}
H = H_{\langle \boldsymbol{a} \rangle} \supseteq \ldots \supseteq 
H_{\langle \boldsymbol{a}_1 \rangle} \oplus \ldots \oplus 
H_{\langle \boldsymbol{a}_n \rangle} \\
\supseteq \Big(H_{\langle \boldsymbol{a}_{11}\rangle}
\oplus \ldots \oplus H_{\langle \boldsymbol{a}_{1q_1}\rangle}\Big)
\oplus \ldots \oplus \Big( H_{\langle \boldsymbol{a}_{n1}\rangle}
\oplus \ldots \oplus 
H_{\langle \boldsymbol{a}_{nq_n} \rangle} \Big) \supseteq \ldots
\end{multline}
in which for every pair of subsequent decompositions, namely $\boldsymbol{a} = \boldsymbol{a}_1 + \ldots + \boldsymbol{a}_n$ and the subsequent one $\boldsymbol{a} 
= \Big(\boldsymbol{a}_{11} + \ldots + \boldsymbol{a}_{1q_1}\Big) + \ldots + 
\Big(\boldsymbol{a}_{n1} + \ldots + \boldsymbol{a}_{nq_n}\Big)$ (where
$\boldsymbol{a}_i = \boldsymbol{a}_{i1} + 
\ldots \boldsymbol{a}_{iq_{i}}$ are the corresponding 
sub decompositions of $\boldsymbol{a}_i$) we 
have the inclusions of the corresponding Hilbert spaces 
\begin{displaymath}
H_{\langle \boldsymbol{a}_i \rangle} \supseteq H_{\langle \boldsymbol{a}_{i1}
\rangle} \oplus H_{\langle \boldsymbol{a}_{i2}
\rangle} \oplus \ldots \oplus H_{\langle \boldsymbol{a}_{iq_i}
\rangle}.
\end{displaymath}
Consider next the intersection $H_{\boldsymbol{a':a}}$ of all 
decreasing Hilbert spaces (\ref{H-net}). It is the immediate analogue of the 
Hilbert space $H_{\boldsymbol{a':a}}$ constructed in Remark of 
Sect. ~\ref{example2}. By construction $\boldsymbol{a}, \boldsymbol{a'}$, as
well as all components $\boldsymbol{a}_i , \boldsymbol{a'}_i$ of the considered decreasing net (sequence) of decompositions belong to $H_{\boldsymbol{a':a}}$. 
Note also that for every Hilbert space (\ref{H-dec}), corresponding to any decomposition $\boldsymbol{a} = \boldsymbol{a}_1 + \ldots + \boldsymbol{a}_n$ 
of our decreasingly directed family (net, we choose it to be a sequence) of decompositions, we have the intersection property
\[
H_{\boldsymbol{a':a}} \bigcap \Big(H_{\langle \boldsymbol{a}_1 \rangle} \oplus \ldots \oplus H_{\langle \boldsymbol{a}_n \rangle} \Big)
= H_1 \oplus \dots \oplus H_n =  H_{\boldsymbol{a':a}} \,\,\,
\textrm{\emph{where}} \,\,\, H_i \subseteq H_{\langle \boldsymbol{a}_i \rangle}.
\]   
Writing symbolically the sequence
of intersections of $H_{\boldsymbol{a':a}}$ (i.e. decompositions of 
$H_{\boldsymbol{a':a}}$) with the sequence of Hilbert
spaces (\ref{H-net}) by
\begin{equation}\label{H-ratio-dec}
H_{\boldsymbol{a':a}} = \ldots = H^{i}_1 \oplus \dots \oplus H^{i}_{n_i} =
H^{i+1}_1 \oplus \dots \oplus H^{i+1}_{n_{i+1}} = \ldots = H_{\boldsymbol{a':a}} 
\end{equation}
we obtain a decreasingly directed net (sequence) of direct sum decompositions
of $H_{\boldsymbol{a':a}}$ in which every direct summand $H^{i}_k$ of a decomposition $H^{i}_1 \oplus \dots \oplus H^{i}_{n_i}$ is equal to
a direct sum $H^{i}_k = H^{i+1}_{k_{1}} \oplus \ldots \oplus H^{i+1}_{k_{p}}$
of some of the direct summands of the subsequent decomposition 
$H^{i+1}_1 \oplus \dots \oplus H^{i+1}_{n_{i+1}}$.

To every such decomposition $H^{i+1}_1 \oplus \dots \oplus H^{i+1}_{n_{i+1}}$
in the sequence (\ref{H-ratio-dec}) of decompositions of $H_{\boldsymbol{a':a}}$, consider the commutative algebra $\mathcal{C}_i$ (as algebra of operators in 
$H_{\boldsymbol{a':a}}$) corresponding to it via the general von Neumann theory of Hilbert space decompositions.  
By construction we obtain in this way an increasing commutative
von Neumann decomposition algebras, which mutually commute. (Note in passing
that all derivations (\ref{deriv-sp}) corresponding to the decompositions of our deceasing family (sequence) of decompositions commute, as well as the corresponding
spectral sum operators (\ref{sp-op}), as a consequence of Lemma III.2 of 
\cite{BI1}.) Consider the weak closure of the sum of all of them, and denote by 
$\mathcal{C} = \Big( \bigcup \mathcal{C}_i \Big)''$. To the algebra $\mathcal{C}$ corresponds a direct integral decomposition
\begin{equation}\label{H-ratio-intdec}
H_{\boldsymbol{a':a}} = \int_{\Sigma} H_\sigma \, d\mu(\sigma) 
\end{equation}
of the Hilbert space $H_{\boldsymbol{a':a}}$ into the Hilbert
spaces $H_\sigma$ with respect to a measure $\mu$ on $\Sigma$. 
This is the immediate analogue of the direct sum decomposition
$H_{\boldsymbol{a':a}} = H_{\langle \boldsymbol{a_1} \rangle} 
\oplus \ldots \oplus H_{\langle \boldsymbol{a}_n \rangle}$
into Hilbert spaces generated by minimal faces of the finite 
dimensional example, compare Remark of Sect. ~\ref{example2}. Any
element of the direct integral Hilbert space (\ref{H-ratio-intdec})
and in particular $\boldsymbol{a}_i \in H_{\boldsymbol{a':a}}$ 
can be written as the vector integral
\[
\boldsymbol{a}_i = \int_{\Sigma} \boldsymbol{a}_{i}(\sigma) \, d\mu(\sigma) 
\] 
of a vector valued function $\sigma \mapsto \boldsymbol{a}_{i}(\sigma)$, 
in which $\boldsymbol{a}_{i}(\sigma) \in H_\sigma$ for almost every
$\sigma \in \Sigma$ and for which 
\[
\int_{\Sigma} |\boldsymbol{a}_{i}(\sigma)|^2 \, d\mu(\sigma) < \infty.
\]
For any two incomparable components $\boldsymbol{a}_i$ and $\boldsymbol{a}_j$
of any decomposition of our decreasingly directed family (sequence)
of decompositions we have 
\[
\operatorname{supp}\boldsymbol{a}_i 
\cap \operatorname{supp}\boldsymbol{a}_j = \emptyset
\]
(up to a $\mu$-measure zero set), where $\operatorname{supp}\boldsymbol{a}_j$ 
and $\operatorname{supp}\boldsymbol{a}_j$ are $\Sigma$-supports of the functions
$\sigma \mapsto \boldsymbol{a}_{i}(\sigma)$
and $\sigma \mapsto \boldsymbol{a}_{j}(\sigma)$. Indeed, by construction 
orthogonal projectors 
$P_1$ and $P_2$ in $H_{\boldsymbol{a':a}}$ on $\boldsymbol{a}_1$ and 
$\boldsymbol{a}_2$ commute and are contained in $\mathcal{C}$ (recall that also all
spectral sum operators (\ref{sp-op}) and derivation operators (\ref{deriv-sp}) corresponding to the decompositions of $\mathcal{D}$ commute in $H$, as a consequence e.g. 
of the lemma III.2 of \cite{BI1}) and thus they commute in $H_{\boldsymbol{a':a}}$ and so with $\mathcal{C}$).

Note also that 
\[
\operatorname{supp}\boldsymbol{a'}_i \subseteq
\operatorname{supp}\boldsymbol{a}_i
\]
for $\boldsymbol{a}_i$ corresponding to $\boldsymbol{a'}_i$.

Assume now, that our family $\mathcal{D}$
 (sequence) of decreasingly directed decompositions 
fulfils the condition 2) of the generalized 
completeness of Subsect. ~\ref{gen-ratio}.
By the condition 2) of the generalized completeness 
 and by the property (\ref{lambda-ineq})
we have
\[
\begin{array}{l}
\boldsymbol{a'} = \int \limits_{\Sigma} f(\sigma) \boldsymbol{a}(\sigma)
 \, d\mu(\sigma) \,\,\, \textrm{\emph{where}} \,\,\, f = \inf f_i , \\
\lambda_1 \boldsymbol{a}_1 + \dots + \lambda_n \boldsymbol{a}_n = 
 \int \limits_{\Sigma} f_i(\sigma) \boldsymbol{a}(\sigma) \, d\mu(\sigma),
\end{array}
\] 
where $f$ and $f_i$ are real valued positive functions of $L^2(\Sigma, \mu)$,
$f_i$ are step-wise functions bounded from below, 
and $\inf f_i$ is with respect to the ordinary order 
in the function space with
$f_i$ corresponding to the $i$-th decomposition 
$\boldsymbol{a} = \boldsymbol{a}_1 + \dots + \boldsymbol{a}_n$ 
of our decreasingly directed 
sequence of decompositions. Thus on account of a well known theorem of measure 
theory $f_i$ converge to $f$ in the $L^2(\Sigma, \mu)$-norm and by the ordinary 
properties of Hilbert space integrals it means that 
$\lambda_1 \boldsymbol{a}_1 + \ldots + \lambda_n \boldsymbol{a}_n$ 
converge in norm to the element $\boldsymbol{a'}$, so that 
the equality must hold $\boldsymbol{a'} = \boldsymbol{a''}$ for the
limit element $\boldsymbol{a''}$ which has been found above. 
Thus for the derivation $\delta$
constructed above we have $\delta \boldsymbol{a} = \boldsymbol{a'}$.
Because $\boldsymbol{a}$ is separating for derivations of the cone 
$S_{\boldsymbol{b}}$ then such derivation, which maps $\boldsymbol{a}$
into $\boldsymbol{a'}$, is determined uniquely. 

Now for any other ratio $\boldsymbol{c':c} = \boldsymbol{a':a}$ we obtain
in this way the same derivation, as immediately follows from the above 
construction of $\delta$ and from the generalized definition of equality
of ratios. Indeed, for the corresponding $\boldsymbol{c}_i$ we have
$\langle \boldsymbol{c}_i \rangle = \langle \boldsymbol{a}_i \rangle$
with the same $\lambda_i$, and thus we will obtain exactly the same sequence of corresponding derivations weakly converging to the same $\delta$.
Theorem ~\ref{twr1} thus follows.

In proving Theorem ~\ref{twr2} we can assume the
derivation $\delta$ to be positive as a selfadjoint operator (positive
spectrum). Consider the increasing spectral family
$\lambda \mapsto \pi(\lambda)$
of the operator $\delta$ analogously as in the proof of Theorem ~\ref{twr2}
for the finite dimensional $H$ of Sect. ~\ref{example2}.
Divide the spectral interval $[0,\parallel \delta \parallel]$
into subintervals $[0,\lambda_1], (\lambda_1, \lambda_2], \ldots (\lambda_{n-1}, \lambda_n ]$ and consider the corresponding spectral projectors 
$P_{\lambda_i - \lambda_{i-1}} = \pi(\lambda_i) - \pi(\lambda_{i-1})$. 
Then by lemma ~\ref{lem} the
subset $F_{\lambda_i} = P_{\lambda_i - \lambda_{i-1}}H \cap 
S_{\boldsymbol{b}}$ is a closed face, compare the proof of Theorem
~\ref{twr2} for finite dimensional $H$ in Sect. ~\ref{example2} (compare also the 
proof of prop. III.1 in \cite{BI1}). To every face $F_{\lambda_i}$ consider
its order unit $\boldsymbol{a}_i$: $\langle \boldsymbol{a}_i \rangle
= F_{\lambda_i}$. By construction because the topological interior of 
$S_{\boldsymbol{b}}$ is non-empty $\boldsymbol{a} = \boldsymbol{a}_1 + 
\ldots + \boldsymbol{a}_n$ must be an order unit separating for derivations.
Let $\boldsymbol{a'} = \delta \boldsymbol{a}$, then we define
the ratio $\boldsymbol{a':a}$ as corresponding to $\delta$. 
Exactly as above we construct the Hilbert space $H_{\boldsymbol{a':a}}$,
using the Hilbert spaces $H_{\langle \boldsymbol{a}_1 \rangle} \oplus 
\ldots \oplus H_{\langle \boldsymbol{a}_n \rangle}$ corresponding to subdivisions $[0,\lambda_1], (\lambda_1, \lambda_2], \ldots (\lambda_{n-1}, \lambda_n ]$
of the spectral interval which compose a decreasingly directed sequence of 
subdivisions. By construction the derivations 
\[
\lambda_1 \delta_{\langle \boldsymbol{a}_1 \rangle} + \ldots 
+ \lambda_n \delta_{\langle \boldsymbol{a}_n \rangle}
\]
converge weakly to $\delta$ (the proof is essentially the same as in the
construction of the proof of Theorem ~\ref{twr2}). 
Next note that if $\boldsymbol{c':c}$ is another ratio corresponding to $\delta$ 
then $H_{\boldsymbol{c':c}} = H_{\boldsymbol{a':a}}$ by the uniqueness
of the spectral family $\pi(\lambda)$ of $\delta$. 
Finally note that for $\boldsymbol{c':c}$ which is comparable with $\boldsymbol{a':a}$,
we also must have $H_{\boldsymbol{c':c}} = H_{\boldsymbol{a':a}}$, because for the
decreasing sequence of decompositions of $\boldsymbol{a}$ there must
exists corresponding sequence of decompositions of $\boldsymbol{c}$
such that for any corresponding components of those decompositions we 
have $\boldsymbol{c'}_i \in \langle \boldsymbol{a}_i \rangle$ 
and $\boldsymbol{a'}_i \in \langle \boldsymbol{c}_i \rangle$ 
and $\langle \boldsymbol{c}_i \rangle = \langle \boldsymbol{a}_i \rangle$.
Because $\delta$ by construction acts as a multiplication (``diagonal'')
operator in $H_{\boldsymbol{c':c}} = H_{\boldsymbol{a':a}}$, then
the two ratios $\boldsymbol{c':c}$ and $\boldsymbol{a':a}$ must be equal
in the sense of the generalized definition.

Summing up
\begin{enumerate}
\item[]
\emph{The Theorems ~\ref{twr1} and ~\ref{twr2} are also true for infinite dimensional
and separable $H$ with a self dual facially homogeneous cone $S_{\boldsymbol{b}}$
with non empty topological interior.} 
\end{enumerate}   

Having the Theorems ~\ref{twr1} and ~\ref{twr2} at hand the assertions 1) and
2) at the beginning of this Section follow now from the results 
presented in  \cite{I}, VI.2, restricted to cone with non empty topological
interior. 

\vspace*{0.5cm}

{\bf Remark 1} \,\,\, Note that the commutative von Neumann algebra
$\mathcal{C}$ is maximal in $H_{\boldsymbol{a':a}}$ iff all the 
Hilbert spaces $H_\sigma$ in (\ref{H-ratio-intdec}) are one dimensional
and thus iff the following theorem holds:
\emph{if dim$\langle \boldsymbol{c}\rangle >1$ then $\langle \boldsymbol{c}\rangle$
is not a minimal face, i.e. $\boldsymbol{c}$ is a sum of two incomparable
positive components.} (Recall that for every closed face $F$
we have $F = \langle \boldsymbol{c}\rangle$ for some $\boldsymbol{c}$.) 
This is the infinite dimensional 
analogue of the property that
every minimal face is one dimensional. In particular if $\mathcal{C}$ 
is maximal in $H_{\boldsymbol{a':a}}$, then all faces corresponding to 
the discrete points of the support of the measure $\mu$ in the formula 
(\ref{H-ratio-intdec}) are one dimensional. If this is not the case, and thus
$H_\sigma$ is higher dimensional for $\sigma$ of a set of positive $\mu$-measure,
then not all elements $\boldsymbol{\eta}$ of the Hilbert space 
$H_{\boldsymbol{a':a}}$ can compose a ratio $\boldsymbol{\eta : a}$ with
$\boldsymbol{a}$, exactly as in the finite dimensional example, compare Remark 
of Sect. ~\ref{example2}. Note also that the inclusions $\supseteq$
of (\ref{H-net}) turn into equalities if and only if the Riesz
additivity postulate is assumed. In this case $\mathcal{C}$ is maximal,
the algebra of ratios is commutative and equal $\mathcal{C}$. 
If this is so, then there exists a set of
positive ``elements'', $\boldsymbol{a}_\sigma$, $\sigma \in \Sigma$ (direct
integral components $\boldsymbol{a}_\sigma = \boldsymbol{a}(\sigma)$ of 
any order unit $\boldsymbol{a} \in H_{\boldsymbol{a':a}}$ viewed as the direct integral 
(\ref{H-ratio-intdec}) of Hilbert spaces $H_\sigma$) whose ``direct integral 
combinations'' span $H$, which in this case is equal to $H_{\boldsymbol{a':a}}$.
Thus any order unit is comparable with any other geometric object, as well as 
all ratios are comparable. This is not true of course if the Riesz additivity
is violated and the inclusions $\supseteq$ in (\ref{H-net}) are proper.  

\vspace*{0.5cm}

{\bf Remark 2} \,\,\, Note that in fact we do not use the so called facial
spectral theorem for derivations of facially homogeneous self dual cones,
but rather we give a proof of it inspired by a generalized theory of ratios, 
independent of that given in \cite{BI1} and repeated in \cite{I};
however, the proof presented here works only for the special case of 
cones with non empty topological interior. The Theorems ~\ref{twr1}
and ~\ref{twr2} are in fact implicitly equivalent to the facial 
spectral theorem for derivations. Let us formulate this theorem 
in full generality as it stands in \cite{BI1}:
\begin{enumerate}
\item[]
\emph{Let $H$ be a real Hilbert space and $H^+$ be a self-dual, facially 
homogeneous (and symmetric) cone in $H$. Then for any selfadjoint derivation
there exists a unique increasing family of faces $\mathbb{R} \ni
\lambda \mapsto F(\lambda)$ such that}
 \[
\begin{array}{l}
\textrm{i)} \,\,\, F(\lambda) = F(\lambda)^{\bot \bot} \,\,\, \textrm{\emph{for all $\lambda$}}.\\
\textrm{ii)} \,\,\, F(a -\epsilon) = F(b + \epsilon)^\bot = 0 \,\,\, 
\textrm{\emph{for all $\epsilon >0$ if spec $\delta \subseteq [a,b]$}}. \\
\textrm{iii)} \,\,\, \bigcap _{\epsilon >0} F(\lambda + \epsilon) = F(\lambda).\\
\textrm{iv)} \,\,\, \delta = \int \limits_{a^-}^{b^+} \, \lambda \,
d \delta_{F(\lambda)} \,\,\, (\textrm{\emph{Lebesque-Stieltjes weak integral}}).
\end{array}
\]
\end{enumerate}

\section{Multiplication of measured quantities according to Newton}
\label{multiplication}

As we have mentioned in Sect. ~\ref{intro} Newton distinguishes between
the geometric quantities which can be measured and the ratios acting on 
them as operations. No natural multiplication can be introduced into the space
of geometric quantities with magnitude, or with dimension, as we would say today, 
which can be measured by ratios. Today when the system(s) of units are 
standardized we could easily overlook this fundamental difference between
dimensionless ratios and measured (geometric) quantities with dimension.
But indeed Newton is right: no natural multiplication can be introduced
into the space of quantities with dimension. However, in the Hellenistic
theory (followed by Newton) multiplication\footnote{As composition of operations}
of ratios $\boldsymbol{a':a}$ and  $\boldsymbol{b':b}$, say of segments 
$\boldsymbol{a'}, \boldsymbol{a},\boldsymbol{b'}, \boldsymbol{b}$ of the line,
can be represented by the ratio $\boldsymbol{a'}\times \boldsymbol{b':}
\boldsymbol{a} \times \boldsymbol{b}$,
where $\boldsymbol{a'}\times \boldsymbol{b'}, \boldsymbol{a}\times \boldsymbol{b}$
are the respective rectangles. Indeed, as we have already explained Newton understood
ratios as operations acting on measured quantities, just like Sumerians
did of fractions -- independently of the representation. Thus the eventual proof 
of this assertion he could realize in the following three steps.
1) The assignment $\boldsymbol{a} \mapsto \boldsymbol{a} \times \boldsymbol{b}$
(fixed $\boldsymbol{b}$) gives an equivalent representation of the algebra of 
ratios acting in the space of rectangles. 2) $\boldsymbol{a':a}$ and  
$\boldsymbol{b':b}$ can be represented by  $\boldsymbol{a'}\times \boldsymbol{b:}
\boldsymbol{a} \times \boldsymbol{b}$ and $\boldsymbol{b'}\times \boldsymbol{a':}
\boldsymbol{b} \times \boldsymbol{a'}$. 3) The assertion follows now from the 
definition of multiplication of ratios as composition of operations. 

This approach cannot be mistaken with the so called 
\emph{geometrical algebra}\footnote{Term used by Heath in his commentaries 
to \emph{Elements} and Archimedes' writings.}
of ancient Greeks, understood as independent of the non-elementary
theory of ratios, which is beset with many limitations: e.g.
no way of comparing rectangles with cubes. 

Newton generalizes this approach, in a way suggested by his experimentally
motivated mathematics. In reality of experimental practice, 
as he notices at the beginning of 
his \emph{Principia}, for any pair of two (geometric) quantities
$\boldsymbol{v}_1$ and $\boldsymbol{v}_2$ with magnitude, which can be measured,
there is another quantity $\boldsymbol{v}_1 \otimes \boldsymbol{v}_2$ 
with magnitude (of another dimension) which
depends linearly on the first $\boldsymbol{v}_1$ and linearly on the second
$\boldsymbol{v}_2$, and as we could say today that it is (or not) universal bilinear 
-- just our tensor product (or its quotient). And it is because the sufficiently 
great supply of such bi-linearly and multi-linearly dependent quantities do exist
that we can introduce a \emph{multiplication} into the space of geometric 
quantities composing together a hierarchy of geometric quantities (we would say today
a tensor product algebra) composed of the two-, three-, and higher-linearly
dependent quantities. Such bi- or multi-linear dependence is called by Newton
\emph{conjunct dependence}. Thus for example if all important observables, or results
of all experiments could ultimately be expressed by quantities of a fixed 
hierarchy of quantities, composing a fixed tensor algebra, then this algebra should play a fundamental role.  
In particular, speaking in a slightly
more modern terms, if all quantities of the hierarchy are magnitudes whose dimensions
are powers of [cm] and [s], which is already suggested by Newton's mechanics
itself, as this is mechanics of motion, and motion may be characterised in 
terms of [cm] and [s], and confirmed by the subsequent physics\footnote{As emphasized
by Einstein, every observation or measurement ultimately rests on the coincidence
of two independent events at the same space-time point. This observation seems to have much more profound justification then Einstein himself could imagine. For example
without the locality principle in quantum field theory we lost physical
interpretation for the scattering processes, not to mention the geometric interpretation
of quantum particle within the Haag's algebraic QFT. This seems to be ignored by those who try to ``quantize gravity'' along the standard lines.}, then the
algebra of space-time should play (and as we already know does) a fundamental role
in physics. Notwithstanding its role (or any other algebra), so profound
or not, it is the sufficiently great supply of multi-linearly depended quantities
which allows us to construct the tensor algebra.    
In this situation there is a distinguished linear space
$V$, preferably finite dimensional which generates the algebra as the
tensor algebra $T(V)$ over $V$.
It is not at all trivial if such a sufficient supply of of quantities $\boldsymbol{v}_i \in V$ and quantities $\boldsymbol{v}_1 \otimes \boldsymbol{v}_2$,
$\boldsymbol{v}_1 \otimes \boldsymbol{v}_2 \otimes \boldsymbol{v}_3$ \ldots multi dependent on $V$ indeed do exist in reality nor if the multi linear dependence
$\otimes$ is symmetric. In the second case the tensor algebra is non commutative.

Before we proceed further on let us quote the first two
definitions of \emph{Principia} together with some fragments of the
comments attached to them (according to Motte \cite{motte}) where Newton
speaks about this:

\vspace{0.5cm}

``\emph{DEFINITION I. The quantity of matter is the measure of the same, arising
from its density and bulk conjunctly.}

\vspace{0.3cm}

\emph{Thus air of a double density, in a double space, is quadruple in quantity; in a triple space, sextuple in quantity.} [...]

\vspace{0.3cm}

\emph{DEFINITION II. The quantity of motion is the measure of the same, arising from the 
velocity and quantity of matter conjunctly.}

\vspace{0.3cm}

\emph{The motion of the whole is the sum of the motions of all the parts; and therefore 
in a body double in quantity, with equal velocity, the motion is double; 
with twice the velocity, it is quadruple.} [...]''
    
Now Newton proposes to apply his generalized theory of ratios to represent
the $\otimes$ by the multiplication of ratios. And thus ratio of his 
\emph{conjunctly dependent} quantities $\boldsymbol{a'} \otimes \boldsymbol{b'}$ and 
$\boldsymbol{a} \otimes \boldsymbol{b}$ 
 should be equal to the composition of the respective ratios 
$\boldsymbol{a':a}$ and $\boldsymbol{b':b}$. This makes sense
also for non symmetric $\otimes$ because the ratios understood as operators
are not necessarily commutative, as explained in the previous Sections. 

Of course in the degenerate case of one dimensional space $V$ of geometric
quantities, i.e. fulfilling the Archimedes-Eudoxus postulate, 
all ``tensors'' $AB \times AB$, $AB \times AB \times AB$, 
\ldots $AB \times\ldots \times AB$, may be represented
by ratios of segments on the line, which we nowadays interpret
by a theorem that $\mathbb{R} \otimes_{\mathbb{R}} \mathbb{R} \simeq \mathbb{R}$,
meaning that the tensor product over reals $V \otimes V$ of one dimensional spaces
$V$ over the reals is again isomorphic to the same one dimensional space $V$. 
But Newton's treatment is much more wise than ours today, even in this one dimensional
case, i.e. in treatment of multiplication of scalar quantities with 
dimension. Today we multiply them  but before
we fix the units (which is not wise and depends on the units,
as any practising physicist knows). Newton's approach 
is universal and independent of the accepted system of units, 
but requires a much advanced mathematical maturity. 

The situation becomes non trivial if the linear space of measured 
(geometric) quantities $V$ (spanned by generators) is higher dimensional 
and, thus, when the Archimedes-Eudoxus postulate is not fulfilled. In this case 
the tensor algebra $T(V)$ may be higher dimensional, and it may even happen
(depending on the relations posed on linear generators, i.e. on the the base
of the linear space of geometric quantities, because $\otimes$ 
may  not be universal, in which case the tensor algebra generated by it is a 
quotient algebra of the true tensor algebra $T(V)$ of $V$) may have infinite
dimension over the reals. In this case if the suggestion of Newton is 
to be realizable the Hilbert space $H$
of the previous Sections must have infinite dimension in order that
the ratios may provide a faithful representation
of $\otimes$, as Newton suggests it. Even more interesting
situation arises when the distinguished space $V$ of geometric 
quantities has not only higher dimension,
but when $\otimes$ is non-symmetric. Still the suggestion of Newton is realistic,
as we explained it in the previous Sections, because the algebra of ratios
may as well be non commutative, so that $\otimes$ a priori can still
be faithfully represented by ratios, as required by Newton.

\vspace*{0.3cm}

{\bf Remark 1} \,\,\, The reader will consult the way in which we build
today algebras over fields at the purely algebraic level, \cite{lan},
compare in particular the construction of quadratic algebras in the algebraic theory of quantum groups in \cite{man}; and then compare it, please, with a very non-trivial way of 
introducing a Hilbert space representation of the generators for
the quantum compact group according to \cite{wor1}
and \cite{wor2}. Of course if the relation posed on $\otimes$
is that it should be symmetric, then the tensor algebra it generates over 
$V$ is nothing else but the ordinary symmetric tensor algebra $S(V)$ 
isomorphic to the ordinary algebra of polynomial functions on $V$, compare
e.g. \cite{lan}.       

\vspace*{0.3cm}

An important moral of this story is that we should expect a distinguished number
of generators to exist, which generate the tensor algebra (say a basis of $V$). 
Here we touch a very delicate point. Only the tensor method of Newton suggests the existence of distinguished generators equal in number to the dimension of 
the spectrum of the tensor algebra viewed as 
an algebra represented by ratios in the Hilbert space $H$ (or $H_{\mathbb{C}}$).
At the pure measure level of the previous Sections
no generators may be distinguished, as the measure structure is completely
''blind'' as to the dimension of the spectrum of (commutative) von Neumann algebra
and doesn't ``feel'' any topological, differential, and geometric invariants. Indeed recall e.g. the theorem (separable $H_{\mathbb{C}}$): \emph{in any weakly closed (commutative) ring $\mathcal{C}$ in $H_{\mathbb{C}}$, there exists a hermitian operator
$A$ which generates this ring, i. e. the smallest weakly closed algebra of operators
containing $A$ equals $\mathcal{C}$.} This is the differential structure
which we need here, which distinguishes the generators. Although topology 
also distinguishes them, it is not so computationally effective and not
so much algebraically accessible and experimentally suggestive as the 
differential one.

Let us note some interesting corollaries which follows from the results
of Sections ~\ref{example1}, ~\ref{example2} and ~\ref{example3}: 
if $\otimes$ is to be represented faithfully by the ratios 
whose action as ratios coincides with the ordinary linear operator
action and in addition $\otimes$ is non-symmetric, then the
complex structure is necessary, and the ratios representing
$\otimes$ compose a subset of the selfadjont part of a von Neumann
algebra of operators acting in a complex Hilbert space $H_{\mathbb{C}}$.
Only if $\otimes$ is symmetric and the corresponding multiplication 
of ratios is commutative the complex structure is not a priori
necessary and a representation of $\otimes$ as selfadjoint
operators in a real Hilbert space $H$ is still possible. Note the analogy with quantum mechanics, where the Heisenberg commutation rules introduce the ``magic''
imaginary $i$. 

\vspace*{0.3cm}

{\bf Remark 2} \,\,\, Although the analogy with quantum mechanics is evident, 
it cannot be understood too naively. In particular in the ordinary 
(nonrealativistic) quantum mechanics
the algebra of observables is irreducibly represented in the Hilbert space
(it is that there are some additional, accidental circumstances, and the so called von Neumann-Stone uniqueness theorem, that we can distinguish uniquely such a representation). In the case of Newton's multiplication the representation
of ratios is far from being irreducible in $H$ (or $H_{\mathbb{C}}$). 
It is perhaps most easily
visible in the case of commutative multiplication of ratios when the Riesz additivity 
is fulfilled, where all Hilbert spaces $H_{\boldsymbol{a':a}}$ corresponding to
ratios $\boldsymbol{a':a}$ are equal to $H$. The algebra of ratios representing $\otimes$
is contained in this case in the algebra $\mathcal{C}$ of Sect. ~\ref{example3}
(compare Remark 1 of that Sect.). By construction $\mathcal{C}$ does not act 
irreducibly in $H$. Of course the mathematical analysis cannot
recover physical underpinnings which are hidden behind the noncommutative 
multiplication of measured quantities. The simplest relation (if any) may be in this: either 1) the algebra of ratios (representing the algebra generated by representors
of $T(V)$) with its representation in $H$ (or $H_{\mathbb{C}}$) is a subalgebra
of the observable algebra with a class of representations induced by the representation of the observable algebra or 2) vice versa. 
However 1) is only possible for the algebra of quantum fields (and not for
the observable algebra of the ordinary quantum mechanics),
where non-trivial selection type construction, like 1), is possible. Indeed
in the algebraic quantum field theory the selection-type  construction 1)
for classical (in the sense: non superposing) quantities, like charges, 
has already been carried on \cite{haag}. The author has undertaken 
investigation of the possibility 1) within a research summarised in \cite{waw}.  
Physics which is behind the non commutativity of multilication would in this case
(i.e. 1)) be deeply  connected with the division of measured quantities
into two classes : superposing and non superposing (say classical). 
But the problem of the connection between Newton's idea and nowadays physics
is for now open.

\vspace*{0.3cm}

Thus Newton's idea consequently realized leads to the assumption
that there should exist a number of generators (corresponding
to basic measurable quantities $\boldsymbol{v}_i$ of $V$) which together
with the quantities  
$\boldsymbol{v}_i \otimes \boldsymbol{v}_j$, \ldots $\boldsymbol{v}_i \otimes
\ldots \otimes \boldsymbol{v}_j$ \emph{conjunctly dependent} on them
should be represented as operators
(ratios) in a Hilbert space $H_{\mathbb{C}}$ (resp. $H$). These generators, 
viewed as operators in $H_{\mathbb{C}}$ (resp. $H$), should be sufficiently smooth in order to be subject to Newton's 
differential calculus and flexible enough to supply a sufficiently 
reach set of representors of measurable quantities. Now as we know
today commutator $A \mapsto [D,A]$ with a fixed operator $D$ is a natural action representing a derivation (fulfilling Leibniz rule)
as acting on operators in $H_{\mathbb{C}}$. ``Smooth operator'' means that it
lies within the domain of all powers of the derivation (being a ratio of geometric representors of fluents which can be expanded in Taylor series).
Connes in his non-commutative geometry \cite{con} interprets
$D$ as the Dirac operator in the general Atiyah-Singer index theorem
 -- a very deep and non-trivial recognition. 
Can the generalized Newton' idea of generalized ratios and his tensor-like
mode of definition of multiplication of measured quantities throw an independent
light on this recognition? For example can we find an independent way
of constructing appropriate $D$, just as we have found an independent proof of the facial spectral theorem for derivations\footnote{These derivations should not be mixed with the derivation of the differential calculus.} at the pure measure level,
inspired by Newton's approach to ratios? We expect we can, at least 
as for some important aspects. This step requires of course a separate and 
extensive examination, which we postpone to another occasion,
and leave as an open problem for now.

\section{Historical Remarks}
\label{history}

It is truly amazing that the ingenious idea of Newton presented here 
has been overlooked both by mathematicians and by physicists.
Unfortunately the original and consequent notation of Newton 
as well as his usage of the therm \emph{ratio}, as presented in the first three Latin editions of \emph{Principia}, has subsequently been
disturbed in all subsequent translations which reflects the fact that the idea of Newton was generally invisible both from the viewpoint of his contemporaries and also for the subsequent generations. The translator's practice in ``smoothing 
opaque and old style of expression'' has frequently fatal consequences: rather
it is frequently that our look at things is stupid and not the style of our fathers opaque. Perhaps the difficulty with the reception of this idea is that 
Newton joined together two great ideas at once:
\emph{tensor product} and \emph{functional analysis}, each of which taken separately is of that kind which requires of us a dramatic effort of will to notice it, but immediately after it is recognized it seems very simple (erroneously recognized as trivial). And thus it is known among mathematicians that Weyl's book on algebraic structures \cite{W3} would be quite sufficient for nowadays mathematics but essentially the only thing which makes the book dated today is the lack of the construction of tensor product space, such as we can find e. g. in the already cited book \cite{lan}, and which was implicitly used by Newton in his \emph{Principia}. Thus in this respect 
\emph{Pricipia} are algebraically more advanced than Weyl's book \cite{W3} 
written in 20th century by a leading mathematician of that century.   
Almost the same situation we have for the second component idea : \emph{functional analysis}. This idea had found first, systematic and unified formulation 
in the famous research book of Banach \cite{Banach}. Newton's idea, almost forgotten, has subsequently been continued by Riemann and Dedekind but only as it concerns the functional analytic point of view.
However the arithmetic construction of reals by Dedekind \cite{Dedekind}, although
inspired by the definition of equality of ratios as presented in Euclid
\emph{Elements}, get lost the duality between the space of geometric
quantities and their ratios viewed as operators acting in the space.
Dedekind's construction of reals has been concentrated only on the arithmetic side.  
His construction, unfortunately devoid of the representation
aspect, has been continued by Kantorovich \cite{Kant}, who applied the algebraic-order
method to investigate the space of continuous functions on the interval
(or more generally on the compact and locally compact Hausdorff spaces). 
This investigations in turn were continued by Fr\'ed\'eric Riesz \cite{riesz1}, \cite{riesz2} and finally by Mark and Selim Krein \cite{krein1} and \cite{krein2}. M. Krein and S. Krein characterised in purely algebraic-order
therms the space of continuous functions on a compact Hausdorff space
in a representation-free manner. Namely they consider a linear semi-ordered space $E$ over the reals satisfying the following axioms: for any $x \in E$ and any $\lambda \in \mathbb{R}$
\begin{enumerate}
\item[]
{\bf Axiom I}.\,\emph{If $x > 0$, then $x \neq 0$.}

\item[]
{\bf Axiom II}.\,\emph{If $x > 0$ and $y > 0$, then $x+ y > 0$.}

\item[]
{\bf Axiom III}.\,\emph{For every element  $x$ there exists an element $x_+ > 0$ (the positive part of $x$), such that $x_+ - x \geq 0$ and  $x' - x_+ \geq 0$
for any $x'$ that verifies two conditions $x' \geq 0$ and  $x' - x \geq 0$.}

\item[]
{\bf Axiom IV}.\,\emph{If $\lambda > 0$ and $x > 0$, then $\lambda x > 0$.}

\item[]
{\bf Axiom V}.\,\emph{There exists in $E$ an element $u > 0$ such that 
for every $x \neq 0$ of $E$ the set of positive numbers $t$ for which $-tu <x < tu$
is non-empty and its greatest lower bound ($\inf$) is different from zero.}

\end{enumerate} 

(Recall that the Axiom III is equivalent to the existence of a minimal decomposition
$x = x_+ - x_-$ ($x_- = (-x)_+$), compare the Remark of Sect. ~\ref{example1}). Then the greatest lower bound of Axiom V they define as a norm $\parallel x \parallel_u$ of $x$,
norms $\parallel \cdot \parallel_u$ for different
choices of $u$ are equivalent. A linear functional $f$ on $E$ they define to be
positive (written $f>0$) if for every $x>0$ we have $f(x)>0$. The method of Riesz
and Kantorovitch easily show that the dual space also verifies the Axioms I-V
with the norm $\parallel f \parallel_u = \sup_{-u<x<u} |f(x)|$ equal $f(u)$
if $f>0$. Then the set $H_u$ of all positive functionals with norm equal 1 is regularly
convex and by the classical Krein-Milman theorem possesses extreme points, 
the totality of which we denote by $S$. Let us call them \emph{pure states}. 
In this situation they proved the following theorem:

\begin{enumerate}
\item[]
\emph{Let the space $E$ be complete in the norm $\parallel \cdot \parallel_u$. 
Then there is in $E$ a uniquely defined commutative operation of multiplication
$E \times E \ni (x,y) \mapsto xy$ which turns $E$ into a ring with unit $u$
and for which $xy >0$ if $x>0$ and $y > 0$. This operation being established,
a necessary and sufficient condition that a functional $f \neq 0$ is a pure
state is that $f(xy) = f(x)f(y)$. The space $E$ is isomorphic to the space
of all continuous functions on the compact Hausdorf space $S$ with the weak
topology. The isomorphism is given by $E \ni x \mapsto \varphi_x \in C(S)$,
where $\varphi_x(f) = f(x)$.}

\end{enumerate}

However this approach, which seems to be initiated by Dedekind, although it could be inspired by Newton himself,
was devoid of the representation aspect, and by this was not well suited
for generalizations covering interesting non commutative algebras. Only
the representation aspect inscribed naturally into the Eudoxus
theory of ratios allows natural such generalizations. Thus it seems
that Newton's great idea had not been drawn to the attention
of the twentieth century mathematicians also
in its aspect concerning functional analytic view
at the theory of ratios. 

The linkage of the representation free approach 
of these works of Riesz, Kanorovich and Krein \cite{krein1} and \cite{krein2} 
to the representation theory
goes through the Riesz representation theorem and 
the Lebesgue measure and integral: indeed  for any
positive functional $f$ on $E$ we get a representation
of $E$ (viewed as algebra) in the Hilbert space
$H = L^2(S,\mu)$, for the measure $\mu$ on $S$ given canonically by the 
functional $f$ (interpreted as Radon integral on $E$)
as multiplication algebra (with action given by point-wise multiplication).  
Thus the fundamental representation aspect, recognized at once by Newton,
has been regained in a roundabout way in the historical development of
mathematics. Gelfand's way of inventing his C*-algebras seems to proceed
from the opposite direction: in recognition that for uniformly closed subalgebras of
operators in Hilbert space one can reject the whole Hilbert space representation
ballast and one can essentially characterise them by their purely algebraic structure
(corresponding to the canonical topological structure in the category of its *-representations, as subsequently interpreted within the non-commutative geometry).
Although this greatly simplify technicalities we must nevertheless remember that
it is the representation aspect which lies at the roots of the profound
duality \emph{geomery-algebra} already recognized by Newton in his generalized 
theory of ratios and his construction of multiplication of measured geometric
quantities and which seems to be fundamental for physics.

%\begin{acknowledgements}
%\end{acknowledgements}

\bibliographystyle{spr-chicago}      % Chicago style, author-year citations
\bibliography{example}   % name your BibTeX data base
\nocite{*}


\end{document}